\documentclass[fleqn,usenatbib]{mnras}

\usepackage{newtxtext,newtxmath}

\usepackage[T1]{fontenc}

\DeclareRobustCommand{\VAN}[3]{#2}
\let\VANthebibliography\thebibliography
\def\thebibliography{\DeclareRobustCommand{\VAN}[3]{##3}\VANthebibliography}

\usepackage{graphicx}	
\usepackage{amsmath}	

\usepackage{multirow}
\usepackage[dvipsnames]{xcolor}

\title[Indirect Measurements of Gas Velocities in the ICM]{Indirect Measurements of Gas Velocities in Galaxy Clusters: Effects of Ellipticity and Cluster Dynamic State}

\author[I. Zhuravleva et al.]{
Irina Zhuravleva,$^{1}$\thanks{E-mail: zhuravleva@astro.uchicago.edu} Mandy C. Chen,$^{1}$ Eugene Churazov,$^{2,3}$ Alexander A. Schekochihin,$^{4,5}$
\newauthor
Congyao Zhang,$^{1}$ Daisuke Nagai$^{6}$ 
\\
$^{1}$~Department of Astronomy and Astrophysics, The University of Chicago, Chicago, IL 60637, USA\\
$^2$~Max Planck Institute for Astrophysics, Karl-Schwarzschild-Str. 1, D-85741 Garching, Germany  \\
$^3$~Space Research Institute (IKI), Profsoyuznaya 84/32, Moscow 117997, Russia \\
$^4$~Rudolf Peierls Centre for Theoretical Physics, University of Oxford, Clarendon Laboratory, Parks Road, Oxford OX1 3PU, UK \\
$^5$~Merton College, Oxford OX1 4JD, UK \\
$^6$~Department of Physics, Yale University, New Haven, CT 06520, USA \\
}

\date{Accepted XXX. Received YYY; in original form ZZZ}

\pubyear{2022}

\begin{document}
\label{firstpage}
\pagerange{\pageref{firstpage}--\pageref{lastpage}}
\maketitle

\begin{abstract}
While awaiting direct velocity measurement of gas motions in the hot intracluster medium, we rely on indirect probes, including gas perturbations in galaxy clusters. Using a sample of $\sim 80$ clusters in different dynamic states from \textsc{Omega500} cosmological simulations, we examine scaling relations between the fluctuation amplitudes of gas density, $\delta\rho/\rho$, pressure, $\delta P/P$, X-ray surface brightness, Sunyaev-Zel’dovich (SZ) y-parameter, and the characteristic Mach number of gas motions, $M_{\rm 1d}$. In relaxed clusters, accounting for halo ellipticities reduces $\delta\rho/\rho$ or $\delta P/P$ by a factor of up to 2 within $r_{500c}$. We confirm a strong linear correlation between $\delta\rho/\rho$ (or $\delta P/P$) and $M_{\rm 1d}$  in relaxed clusters, with the proportionality coefficient $\eta \approx 1$. For unrelaxed clusters, the correlation is less strong and has a larger $\eta\approx 1.3\pm 0.5$ ($1.5\pm0.5$) for $\delta\rho/\rho$ ($\delta P/P$). Examination of the power-law scaling of $M_{\rm 1d}$ with $\delta\rho/\rho$ shows that it is almost linear for relaxed clusters, while for the unrelaxed ones, it is closer to $\delta\rho/\rho\propto M_{\rm 1d}^2$, supporting an increasing role of non-linear terms and compressive modes. In agreement with previous studies, we observe a strong correlation of $M_{\rm 1d}$ with radius. Correcting for these correlations leaves a residual scatter in  $M_{\rm 1d}$ of $\sim 4 (7)$ per cent for relaxed (perturbed) clusters. 
Hydrostatic mass bias correlates with $M_{\rm 1d}$ as strongly as with $\delta\rho/\rho$ in relaxed clusters. The residual scatters after correcting for derived trends is  $\sim 6-7$ per cent. These predictions can be verified with existing X-ray and SZ observations of galaxy clusters combined with forthcoming velocity measurements with X-ray microcalorimeters.
\end{abstract}




\section{Introduction}

The intracluster medium (ICM), filled with hot ($T\sim 10^7 - 10^8$ K), X-ray-emitting gas, is continuously perturbed by matter accretion along cosmic filaments, mergers, feedback processes, motions of galaxies, etc. These processes generate gas motions inside the cluster potential well that contribute significantly to the energy and pressure budget of the ICM and prevent clusters from reaching a full hydrostatic equilibrium \citep[e.g.,][]{Ras06,Nag07a,Iap08,Lau09,Vaz09,Nelson2012,Bat12,Zhuravleva2013,Nelson2014,Shi16,Bif16,Angelinelli20,Barnes21}. Gas motions transfer energy from large to small scales and eventually  dissipate into heat, regulating cooling and star-forming processes, re-accelerate relativistic particles, reorder and amplify magnetic fields in the ICM \citep[e.g.,][]{Sch06,Bru07,Vaz09,Zhu14b,Min15,Shi18,Shi20}. With a few exceptions \citep[e.g.,][]{San10,Tam14,Hit18,San20}, direct velocity measurements await future high-resolution X-ray spectrometers, such as {\it Resolve} on {\it XRISM} (launch 2023, \citealt{xrism20}) and projected missions like {\it Athena} \citep{athena13}, {\it LEM} \footnote{http://lem.cfa.harvard.edu} and $Lynx$ \footnote{https://www.lynxobservatory.com}. Currently, we rely on indirect probes of gas motions, including resonant scattering \citep[e.g.,][]{Gil87,Chu04,Ogo17,HitRS}, 
an impact of gas motions on metal abundance profiles \citep{Reb05}, widths of shocks \citep{Nul13}, and gas density and pressure fluctuations \citep[e.g.,][]{Sch04,Chu12,Zhu14b,Zhu18,Wal15,Kha16}. 

The latter method is particularly attractive as it (1) provides a way of measuring not only velocity amplitudes but also their length scales and power spectra - characteristics that are otherwise difficult to extract from observations; and (2) utilizes high-resolution X-ray imaging data that has been extensively collected with \textit{Chandra} and \textit{XMM-Newton} over the past 20+ years. It is also possible to use sub-mm/SZ images of galaxy clusters to measure velocities through imprinted pressure fluctuations \citep[see][for a review]{Mroczkowski2019}. However, with the limited spatial resolution of {\it Planck} data, only large-scale fluctuations could be constrained \citep{Kha16}.

The idea of the method, employing that density fluctuations (revealed by X-ray images) can be linked to slow gas motions, rests on the consideration of a weakly perturbed stably stratified atmosphere. Namely, it has been shown that in atmospheres of relaxed galaxy clusters with predominantly subsonic motions driven on large, buoyancy-dominated scales \citep{Nag13,Shi18,Shi19,Shi20}, there is a linear relation between the amplitude of density fluctuations, $(\delta\rho/\rho)_{\rm k}$, measured at a wavenumber\footnote{Here, we adopt a wavenumber $k$ that is related to a length scale $l$ without a factor $2\pi$.} $k=1/l$, and a one-component velocity of gas motions, $v_{\rm 1d,k}$, namely, $(\delta\rho/\rho)^2_{\rm k} \approx \eta_{\rm k}^2(v^2_{\rm 1d,k}/c^2_{\rm s})$, where $c_{\rm s}$ is the sound speed within the gas and $\eta_{\rm k}$ is the proportionality coefficient \citep{Zhuravleva2014}. In essence, it is the radial entropy gradient of the atmosphere that sources density variations of displaced gas lumps. In the same paper, the proportionality coefficient was calibrated using a small sample of relaxed and quasi-spherical galaxy clusters from \textsc{Omega500} cosmological simulations \citep{Nag07a,Nag07b,Nelson2014}, giving $\eta_{\rm k} \approx 1.0\pm0.3$. Performing high-resolution 3D hydrodynamic simulations of idealized turbulence in a stratified cluster atmosphere, \citet{Gas14} confirmed this scaling and additionally showed that (1) the scaling may break on small scales in the presence of strong thermal conduction, providing a promising method to constrain conduction in the bulk ICM, and (2) pressure perturbations become substantial when turbulence Mach number $\gtrsim 0.5$, providing an additional source of correlation between velocity and pressure/density fluctuations. 

In soft X-rays (e.g., $0.5 - 3$ keV band for $T>3$ keV gas), the X-ray emissivity per unit volume is proportional to squared gas density in galaxy clusters. Hence, power spectra of density fluctuations can be directly measured from X-ray images. Pressure fluctuation could be probed using  harder X-ray images \citep[e.g.,][see their fig.2]{Forman17} or SZ maps. Measuring velocity power spectra through density fluctuations in a sample of cool cores in relaxed clusters inferred typical velocities between $\sim 100-150$ km/s on scales $< 50$ kpc, and up to $\sim 300$ km/s on larger scales $\sim 100$ kpc \citep{Zhu18}. These velocities are consistent with those measured with \textit{Hitomi} in the core of the Perseus cluster given the dominant scale of motions in this central region is $< 100$ kpc (confirmed with observations, \citealt{Hit18}). The same (or similar) ideas have been extended outside relaxed cool cores, to unrelaxed clusters (including rather extreme cases) and on large spatial scales up to $500-700$ kpc \citep[e.g.,][]{Chu12,Gu09,Hof16,Eck17,Bon18}. 

While these results are encouraging, the velocity - density (or pressure) fluctuations scaling has not been verified for unrelaxed clusters and beyond the inner regions, where the amplitudes of perturbations become large invalidating the assumption of a nearly hydrostatic atmosphere. To explore this problem, a large sample of unrelaxed clusters is required as the expected scatter could be substantially larger than for relaxed systems. This brings us to the main scope of this work to calibrate further  the velocity-fluctuations relation, accounting for different cluster dynamic states, halo ellipticities, and using a large sample of clusters from cosmological simulations. We will consider both density and pressure fluctuations, as well as their observable (``projected'') characteristics, namely X-ray surface brightness, $I_{\rm X}$, and SZ $Y_{\rm SZ}$ parameter. A closely related question, namely, whether this link between velocity and the amplitude of fluctuations is more fundamental (i.e., causal) compared to the one arising from radial trends of increasing levels of perturbations and velocities of gas motions found in cosmological simulations \citep[e.g.,][]{Lau09,Nelson2012,Nelson2014b,Bif16,Vaz17}, will also be examined.

In the past couple of years, there has been an interesting theoretical development in the field. Conducting high-resolution hydrodynamic simulations of subsonic turbulence with different levels of stratification characterized through the Richardson number, $Ri$, \citet{Moh20} showed that the amplitude of density fluctuations, characterized through the standard deviation of the logarithmic density fluctuations, increases with increasing $Ri$ (i.e., with the level of stratification). They further verified the relation between the amplitude of density fluctuations and velocity, showing that it was reaching earlier predictions by \citet{Zhu14b} in the limit of low Mach number and $Ri\gtrsim 10$. They also showed that pressure fluctuations are independent of the level of stratification  and only depend on the Mach number of turbulence \citep[see also][]{Moh21}. Another recent study by \citet{Sim22} explored the density fluctuations-velocity scaling using a sample of 20 clusters (in various dynamic states) from the Itasca cluster sample from cosmological simulations, confirming a linear relation between the amplitude of density fluctuations (namely, the root mean square of density fluctuations) and velocity. They found that relaxed objects show a slightly steeper slope $\eta=1.1\pm0.06$ compared to the earlier predictions, which is consistent with $1\pm 0.3$ given a large scatter. Perturbed clusters followed a flatter and weaker relation. In contrast to \citet{Moh20}, they did not find a strong correlation between the logarithmic density fluctuations and the level of stratification. 

Our study further explores the scaling relations between various amplitudes of fluctuations and velocities in clusters in different dynamic states, accounting for radial variations of these characteristics and halo ellipticity. Besides a fundamental interest in gas dynamics in the ICM, measuring velocities of gas motions and associated non-thermal pressure is important for precise cluster mass measurements through hydrostatic equilibrium for cosmology \citep[e.g.,][for a recent review]{Pratt2019}. 
Given a link between the amplitudes of fluctuations and velocities, we examine these amplitudes as potential proxies for the cluster mass bias. 

The structure of the paper is as follows. Simulations used for this work, sample selection, and classification of clusters are described in Section \ref{sec:sample}. Methodology, in particular, the ellipticity-measurement algorithm, characterization of fluctuations and gas velocity field are summarized in Section \ref{sec:method}. Section \ref{sec:results} shows our main results on fluctuation amplitudes, correlations between the amplitudes and velocities, and calibration of the proportionally coefficient $\eta$ in various cases. A possible connection of fluctuations to cluster mass bias is discussed in Section \ref{sec:discussion}. The main conclusions are summarized in Section \ref{sec:conclusions}.

\begin{figure}
    \centering
    \includegraphics[trim=10 30 0 0, width=0.49\textwidth]{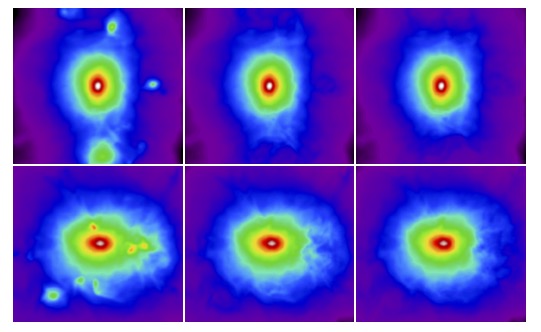}
    \caption{Illustration of the clump removal procedure used to prepare data for the fluctuation analysis. The top and bottom rows show two different clusters (relaxed and unrelaxed, respectively). The left panels show initial maps of projected gas density, while the corresponding middle and right panels show the same projected density after applying the clump removal algorithm with $f_{\rm cut}=3.5$ (middle) and $f_{\rm cut}=2.5$ (right). The box size used for these projected maps is $\sim 5$ Mpc on each side. The color scales of all maps  are the same. The images are lightly smoothed for display purposes. The clumps are effectively removed regardless of their size, density, and location within the cluster.}
    \label{fig:clump_removal}
\end{figure}

\section{Sample of galaxy clusters}
\label{sec:sample}
Our sample includes 78 galaxy clusters from the non-radiative hydrodynamic cosmological simulations \textsc{Omega500} \citep{Nag07a,Nag07b,Nelson2014}. The input cosmology corresponds to a flat $\Lambda$CDM model with $\Omega_m=0.27$, $\Omega_b = 0.0469$, $h=0.7$ and $\sigma_8=0.82$. These simulations use an Adaptive Refinement Tree (ART) hydrodynamic solver \citep{Kra99,Rudd08}, which is particularly good at capturing turbulence, shocks, and sharp contact discontinuities. The default simulation volume is resolved using six levels of mesh refinement, implying a maximum comoving spatial resolution of $\sim 15$ $h^{-1}$ kpc. We checked that a finer refinement of the data does not affect our results. The total masses of clusters in our sample range between $M_{\rm 200c}=2 \cdot 10^{14}h^{-1}M_{\odot}$ and $1.6 \cdot 10^{15}h^{-1}M_{\odot}$ with the median value $\sim 5\cdot 10^{14}h^{-1}M_{\odot}$ at redshift $z=0$. The center of each cluster corresponds to the location of the dark matter particle with the most bounded gravitational energy, which translates to the densest peak of the most massive cluster in cases of merging clusters.

Dense clumps of gas that could be associated with infalling structures or formed in simulations due to incomplete (sub-grid) physics prescription or insufficient resolution could bias mean thermodynamic characteristics of the hot gas in clusters \citep[e.g.,][]{Nag11,Ras14,Avestruz2014}, affect ellipticity measurements (Section \ref{sec:method_ellipticity}), and the amplitude of projected fluctuations and velocity in the bulk gas  (Section \ref{sec:amplitude}). It is crucial to remove them carefully before analyzing fluctuations in the bulk gas\footnote{In observations, such dense clumps are usually identified in X-ray images and removed from the analysis of gas fluctuations.}. Following the method proposed by \cite{Zhuravleva2013}, we identified the clumps through high-density tails of the probability density distribution of gas density within considered regions (see their Fig. 2), using $f_{\rm cut}=3.5$. Namely, in each region, all cells with a density larger than the median density by 3.5 $\sigma$, where $\sigma$ is the log$_{10}$-based standard deviation of density distributions, are associated with the dense clumps. These identified clumps were then removed from the data, leaving the bulk gas component intact. To illustrate the procedure, Fig. \ref{fig:clump_removal} shows two examples of clusters (a relaxed one on top and unrelaxed on the bottom) with prominent clumps. Their initial projected density with all the clumps is shown on the left, while the same projected density with removed clumps is shown in the middle ($f_{\rm cut}=3.5$, default) and right ($f_{\rm cut}=2.5$) panels. The procedure works well and the difference between the two choices of $f_{\rm cut}$ is minor. We checked that varying $f_{\rm cut}$ within a reasonable range,  by $\pm 30$ per cent, does not affect our main conclusions.    

\begin{figure}
    \centering
    \includegraphics[trim=10 25 10 0, width=8.5cm]{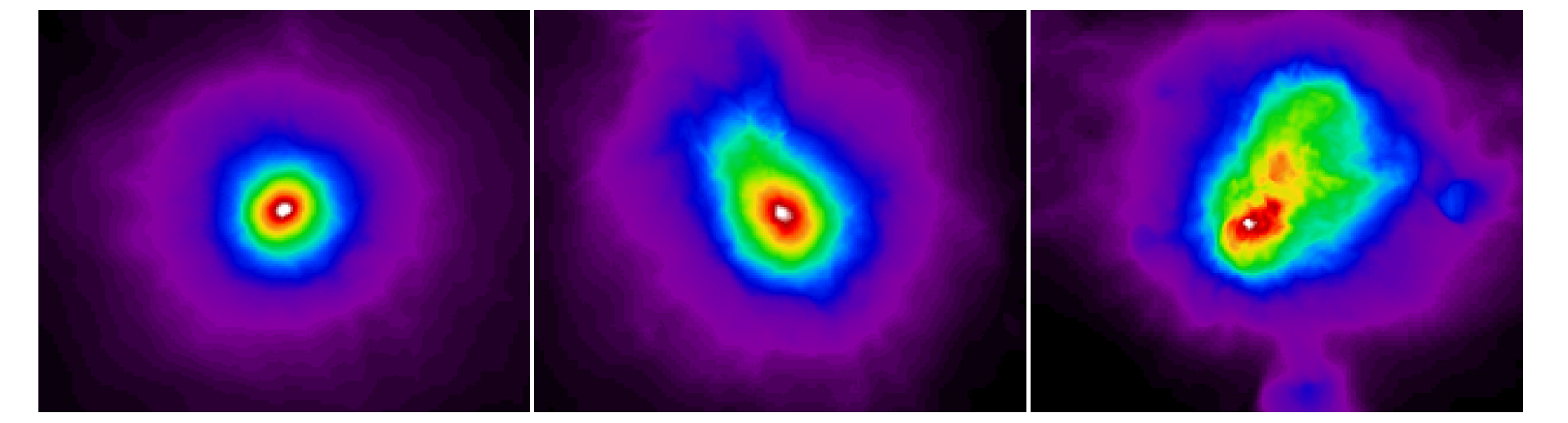}
    \caption{Projected gas density distribution ($\propto \int \rho dl$) of a typical relaxed (left), unrelaxed (right), and in-between (middle) cluster in our sample. White/purple color traces the highest/lowest projected density values. Only data within the radius $\sim 2r_{\rm 500c}$ is used for these images. 
    }
    \label{fig:subsamples}
\end{figure}

Visually inspecting all clusters in our sample, we classified them into three groups: relaxed (smooth morphology), unrelaxed (very perturbed), and in-between \citep[see also][for other classifications]{Shi16,Chen2019}. For the classification, we projected the 3D density distributions (with removed high-density clumps) along three axes and ranked each projection based on how easily the cluster center could be identified (e.g., a few central densest pixels vs. a more diffuse region), the symmetry of gas distribution, the presence of merger-driven structures (e.g., filaments, large-scale contact discontinuities), and how substantial is the clump removal. Fig. \ref{fig:subsamples} shows representative examples of clusters from each group. The final subsamples of relaxed, unrelaxed, and in-between clusters include 19, 27, and 32 objects, respectively. 

Our mass-limited sample is cosmologically representative in terms of cluster dynamical state. This means that the split between relaxed, in-between, and unrelaxed clusters should be roughly similar to samples of nearby clusters. While the visual classification of clusters in these three categories is subjective (although it could be implemented with modern machine-learning techniques), the most important are the fractions, namely 0.24, 0.35, and 0.41 for relaxed, in-between and unrelaxed clusters, respectively.  When using other criteria for apparent deviations from the relaxed state, one can hope that choosing $\sim$25 per cent of the most relaxed clusters would approximately match our relaxed sample.

\section{Methodology}
\label{sec:method}

\subsection{Ellipticity measurements \label{sec:method_ellipticity}}
Gas distribution in the ICM often deviates from perfect spherical symmetry. Cosmological simulations show that the average ellipticity of the hot gas varies depending on the distance from the cluster center and the physics involved in the simulations, reflecting elongation of the underlying gravitational potential dominated by dark matter \citep[e.g.,][]{Lau12,Chen2019,Har21}. 
Using spherical shells to characterize the amplitude of fluctuations in elongated clusters could bias the amplitude, effectively increasing it on large scales. Therefore, we performed the analysis in both spherical and elliptical shells. 

After removing high-density clumps, we replaced the removed substructure with the median values of considered characteristics at that cluster-centric radius and estimated the ellipticity of ICM at each radius using an iterative method described in \cite{Zemp2011}. Summarizing the method,  we first calculate the shape tensor in each spherical shell as
\begin{equation}
    S_{ij} = \frac{\sum_k \rho_k (\mathbf{r}_k)_i (\mathbf{r}_k)_j}{\sum_k \rho_k} \, ,
     \label{eq:ell}
\end{equation}
where $i$, $j$ indices represent the $x$, $y$ or $z$ axes, $({\bf r}_k)_j$ is the $j-$component of the position vector of the $k$th cell, and $\rho_k$ is the gas density in the $k$th cell. The eigenvalue of $S_{ij}$ estimates the axis ratio of the ellipse in the $ij$ plane, and the eigenvector represents the orientation of the ellipse. We then update the $S_{ij}$ tensor with a new shell definition that takes into account the ellipticity and orientation of the ellipsoid and repeat the calculation until it converges. For the ellipticity measurements of gas pressure distribution, we follow the same procedure, substituting density with the gas thermal pressure in relation (\ref{eq:ell}).

Note that for an ellipsoidal shell, the definition of the radius is given by 
\begin{equation}
    r_\mathrm{ell} = \sqrt{x_\mathrm{ell}^2+\frac{y_\mathrm{ell}^2}{(b/a)^2}+\frac{z_\mathrm{ell}^2}{(c/a)^2}} \, ,
\end{equation}
where $x_\mathrm{ell}$, $y_\mathrm{ell}$ and $z_\mathrm{ell}$ are the coordinates along the eigenvectors of the shape tensor $S_{ij}$ after convergence, $a$, $b$ and $c$ are the semi-principal axes with $a\ge b\ge c$. Below, we denote radius as $r$, which corresponds to the semi-major axis $a$ of the ellipsoid surface in elliptical shells.

\begin{figure}
    \centering
    \includegraphics[trim= 0 50 0 50, width=0.5\textwidth]{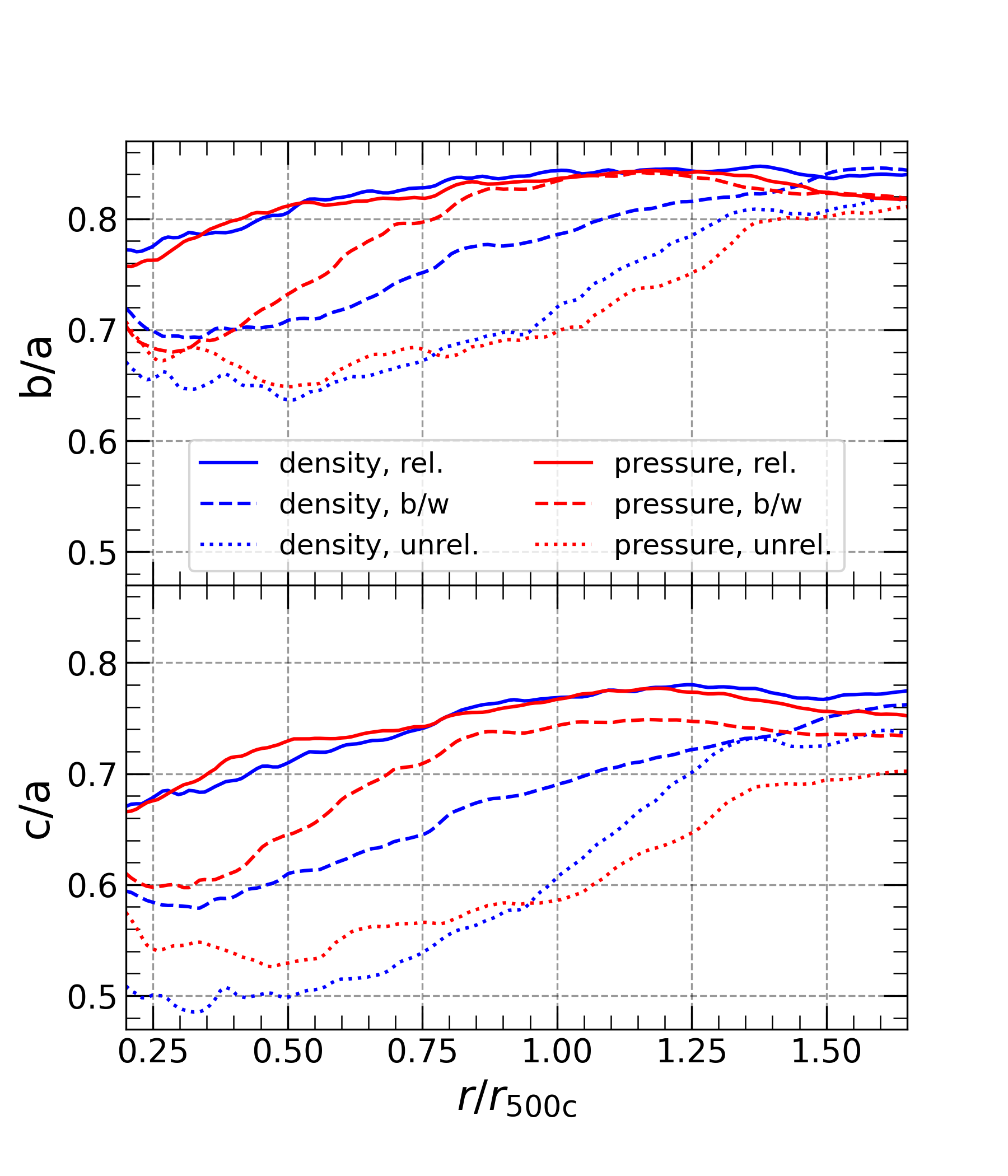}
    \caption{Radial profiles of axis ratios averaged over our subsamples of relaxed, unrelaxed, and in-between clusters. $a, b$ and $c$ are semi-principal axes with $a\ge b\ge c$. The ratios are shown for density and pressure distributions with removed high-density clumps assuming $f_{\rm cut}=3.5$.}
    \label{fig:ellipticity}
\end{figure}

Fig. \ref{fig:ellipticity} shows radial profiles of axis ratios averaged over our samples of relaxed, unrelaxed, and in-between clusters. One can see that relaxed clusters are more spherical compared to unrelaxed ones, especially at $r<r_{500c}$. Ellipticities calculated from density and pressure perturbations are very similar for the relaxed and unrelaxed clusters, however, there are some differences in the in-between group. In all cases, the effects of ellipticity are the strongest within the inner $r_{500c}$ region. Note that if gaps from the high-density clumps are smoothly filled with the density values that separate clumps from the bulk gas (i.e., there are no sharp edges associated with removed clumps), the ellipticity changes maximum by 5 and 7 per cent in relaxed in unrelaxed clusters, respectively. This difference is noticeable beyond $1.5r_{500c}$.

Besides the 3D characteristics, we also consider observational (``projected'') characteristics such as the X-ray surface brightness $I_{\rm X}\propto \int n_{\rm e}^2\Lambda(T)dl$ and SZ $y$-parameter $Y_{\rm SZ}\propto \int n_{\rm e}Tdl$. Here, $n_{\rm e}$ is the electron number density, $\Lambda(T)$ is the X-ray emissivity calculated within the $0.5-2.$ keV band, $T$ is the electron temperature, and the integration is along a line of sight $l$ (the length of the line of sight is $\sim 5.5$ Mpc in our analysis). Elliptical annuli for these observational characteristics are calculated similarly to the 3D case. We calculate the shape tensor (\ref{eq:ell}), using $I_{\rm X}$ or $Y_{\rm SZ}$ instead of density and $i, j$ representing only $x$ and $y$ axes. The eigenvalue and eigenvector of the shape tensor estimate the ellipticity and orientation of the best-fitting ellipse at each distance from the cluster center. Based on the calculated ellipticity, we construct the elliptical projected radial grid used in this work.

\subsection{Characterizing gas fluctuations  \label{sec:method_calculate_fluctuations}}

We calculated the amplitude of the bulk density, pressure, X-ray surface brightness, and SZ $y$-parameter fluctuations (hereafter, we refer to any of these amplitudes as $\delta\xi/\xi$) following the procedure proposed by \citet{Zhuravleva2013}. Consistent with earlier studies, we confirmed that the probability density distributions of these characteristics in each considered region follow log-normal distributions \citep{Kaw07}. In each shell/annulus (spherical or elliptical), we measured the width of the probability density distribution of $\xi$ as $\delta\xi/\xi=\log_{10}({\xi_1/\xi_2})$, where $\xi_1$ is the 12th-quantile of the characteristic's distribution within the shell/annulus, and $\xi_2$ is the 88th-quantile. If the distribution is approximated as a log-normal distribution with a standard deviation (natural log based) $\sigma$, then $\displaystyle  {\frac{\delta\xi}{\xi}=\frac{2\sqrt{2\ln 2}}{\ln 10}\sigma\simeq 1.02\sigma}$. Note that this definition for the total amplitude (i.e., measured on all scales together) of density or pressure fluctuations is not sensitive to the presence of high-density clumps or the procedure used to identify and remove them from the data. 

For density fluctuations, we also considered an average amplitude of fluctuations as a function of an isotropic wavenumber $k=\sqrt{k_x^2+k_y^2+k_z^2}$. This scale-dependent amplitude of density fluctuations has been measured in several observed galaxy clusters using X-ray imaging data \cite[e.g.,][]{Chu12,Wal15,Are16,Zhu18}. Following the observational procedure, we calculated a radial profile of X-ray surface brightness and approximated it with a $\beta-$model for each cluster. Dividing the gas density distribution by the corresponding best-fitting $\beta-$model, we obtained a data cube of density fluctuations. We then calculated a power spectrum of density fluctuations, $P(k)$, using the modified $\Delta-$variance method \citep{Are12,Oss08}. Note that this method of calculating power spectra is not sensitive to gaps in the data (caused by the removal of high-density clumps) and non-periodic boundaries of the data cube. Finally, we calculated the amplitude of density fluctuations, $(\delta\rho/\rho)_k$, at each $k$ as $(\delta\rho/\rho)_k=\sqrt{P(k)4\pi k^3}$. Velocity power spectra were calculated for each velocity component using the same $\Delta-$variance method.   

\subsection{Characterizing gas velocity field}

After removing the high-density clumps from our data cubes, we calculated the characteristic RMS velocity amplitude of the bulk-gas component in each shell/annulus as
\begin{equation}\label{eq:v_amplitude}
    V_{\rm rms} = \sqrt{\langle (V_{\rm x}-\langle V_{\rm x}\rangle)^2+(V_{\rm y}-\langle V_{\rm y}\rangle)^2+(V_{\rm z}-\langle V_{\rm z}\rangle)^2\rangle} \, ,
\end{equation}
where $\langle\rangle$ denotes averaging over all particles within a shell/annulus, and $\langle V_{\rm x}\rangle$, $\langle V_{\rm y}\rangle$, $\langle V_{\rm z}\rangle$ are the components of mean velocities within each region (the reference velocity). Our experiments with different choices of the reference velocity (e.g., averaging within some central regions instead) showed no significant differences in the $V_{\rm rms}$ averaged over the subsamples of clusters. Consistent with previous studies, we saw that the RMS velocity of the bulk component had very regular behavior with radius in contrast to the velocity of the high-density clumps.

It is important to emphasize that we do not decompose gas motions into bulk, laminar motions and genuine turbulence, and consider the whole velocity field in our study. This is different from the other recent study by \citet{Sim22}. Such velocity decomposition relies on an assumption about the filtering scale that is difficult to define unambiguously given the variety of velocity driver scales in the ICM \citep[e.g.,][]{Vaz17}. Moreover, the Reynolds number in cosmological simulations is relatively small, typically below 100. Therefore, the filtered small-scale motions do not necessarily correspond to fully-developed turbulence. Given these difficulties, we choose a conservative approach to explore the whole velocity field. Idealized numerical simulations of merging clusters and turbulence in stratified atmospheres are better suited for addressing the relation between fluctuations and turbulent motions of the ICM gas \citep[e.g.,][]{Gas14,Moh20}.   

We define the characteristic Mach number of gas motions as $M=V_{\rm rms}/c_{\rm s}$, where the sound speed of the gas is $c_{\rm s} =  \sqrt{\gamma k_{\rm B} T/\mu m_\mathrm{p}}$, $\gamma=5/3$ is the adiabatic index for ideal monatomic gas, $k_{\rm B}$ is the Boltzmann constant, $\mu=0.588$ is the mean atomic weight and $m_\mathrm{p}$ is the proton mass. One-component Mach number is formally defined as $M_{\rm 1d}=M/\sqrt{3}$. When exploring projected fluctuations ($y-$parameter or X-ray surface brightness) and their relation to the Mach number of gas motions, we used velocities and sound speeds averaged within shells even if the fluctuations are probed within geometrically-different regions (along the line of sight at a given distance of each annulus from the cluster center). This choice is motivated by our desire to extract radial velocity information in observed clusters. In other words, we are trying to link the potential observables at a given projected distance from the cluster center to gas properties at a similar distance in 3D.

\section{Results}
\label{sec:results}

\subsection{Amplitude of gas fluctuations in relaxed and dynamically perturbed clusters}
\label{sec:amplitude}

\begin{figure*}
    \centering
    \includegraphics[width=\linewidth, trim=0 40 0 0]{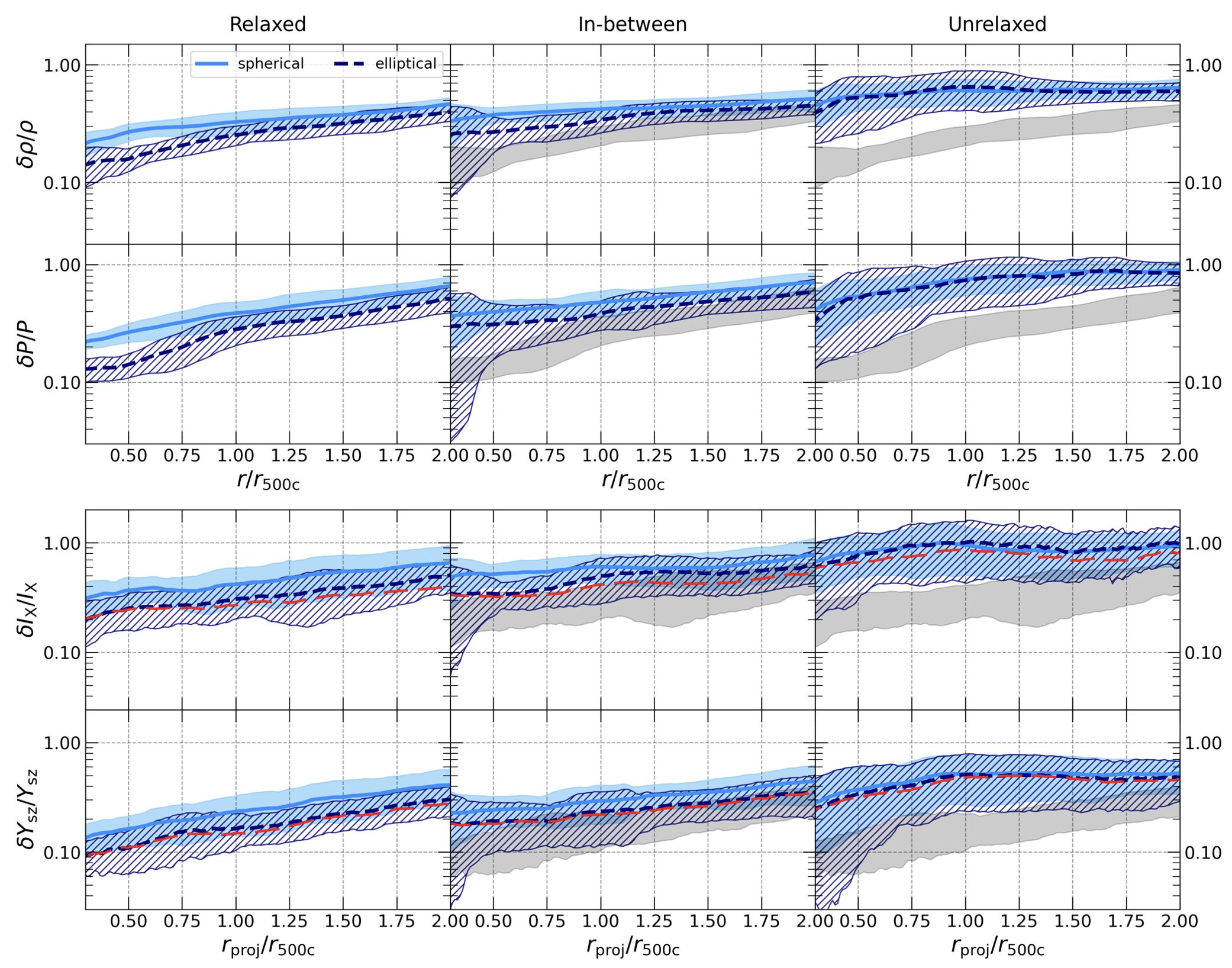}
    \caption{Radial profiles of the amplitudes of density and pressure fluctuations (top two rows) and fluctuations of X-ray surface brightness and SZ $y$-parameter (bottom two rows) averaged over the subsamples of relaxed (left), in-between (middle), and unrelaxed (right) clusters. Light blue and solid curves/regions show results in spherical shells or annuli, while navy dashed/hatched curves/regions - are in elliptical regions. Gray regions show results for relaxed clusters in elliptical shells/annuli and are plotted for visual comparison with clusters in other dynamic states. Dashed red curves show results for $f_{\rm cut}=2.5$ (cf. the navy dashed curves).}
    \label{fig:amplitudes}
\end{figure*}

\begin{figure*}
    \centering
    \includegraphics[width=\linewidth, trim=100 220 100 200]{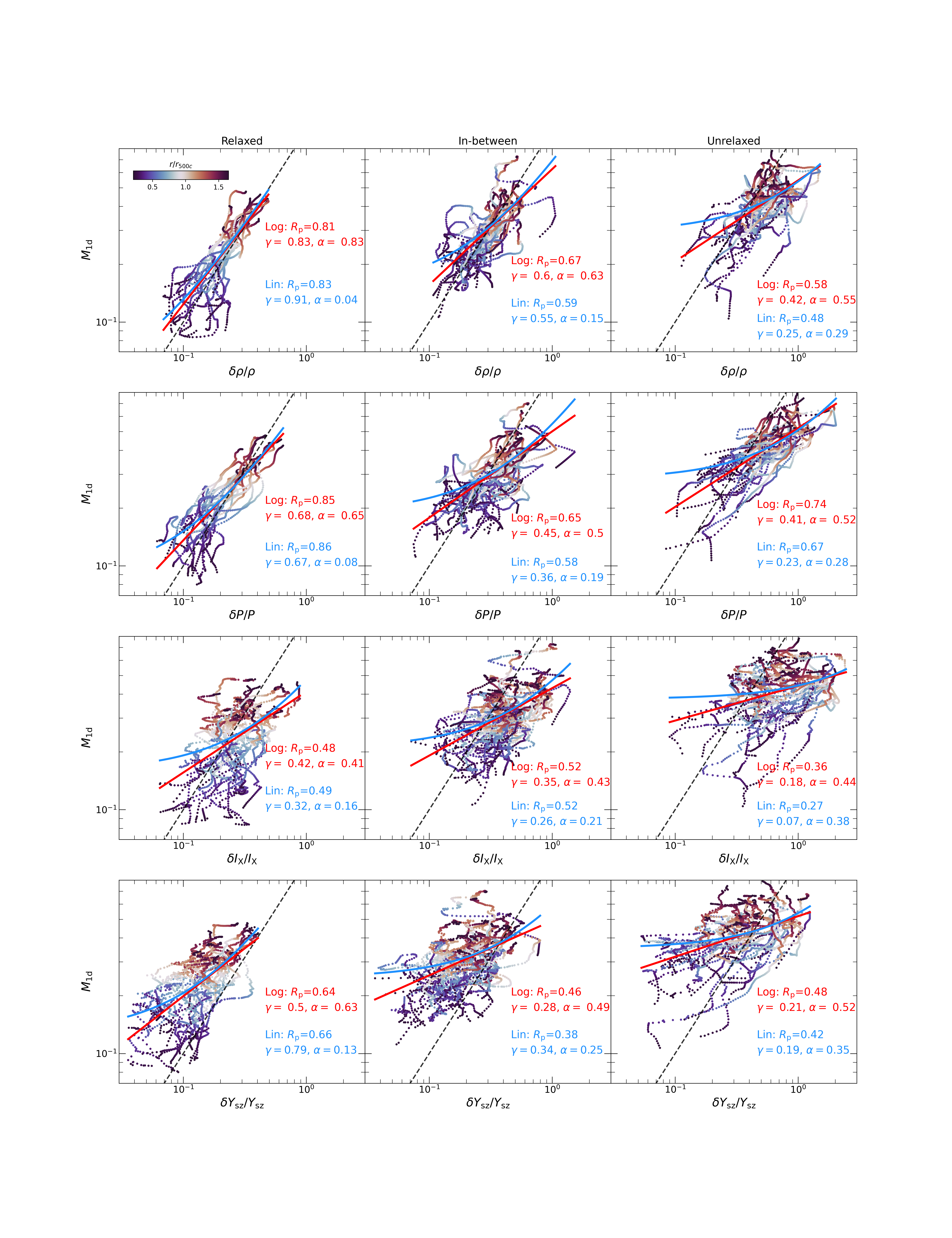}
    \caption{Correlations between the amplitudes of gas fluctuations (from top to bottom: density, pressure, X-ray surface brightness, SZ $y-$parameter) and one-component Mach number of gas motions. Color represents a radial distance from the cluster center. At each radius, fluctuations are calculated within a corresponding elliptical shell/annulus. Each track of points corresponds to an individual cluster. Only measurements within $0.2r_{500c}<r<1.7r_{500c}$ are used. All distributions are fitted with a linear function (blue), $M_{\rm 1d}=\alpha+\gamma(\delta\xi/\xi)$, as well linear function between the log-10 of these characteristics (red), i.e., $M_{\rm 1d}=10^{\alpha_1}(\delta\xi/\xi)^{\gamma}=\alpha(\delta\xi/\xi)^{\gamma}$. Corresponding Pearson's ranks, $R_{\rm p}$, and the best-fitting parameters are shown in legends. The dashed gray lines indicate a one-to-one correlation for visual guidance. See Section \ref{sec:calib} for details.}
    \label{fig:corr_all}
\end{figure*}

\begin{figure*}
    \centering
    \includegraphics[width=\linewidth, trim=15 30 0 0]{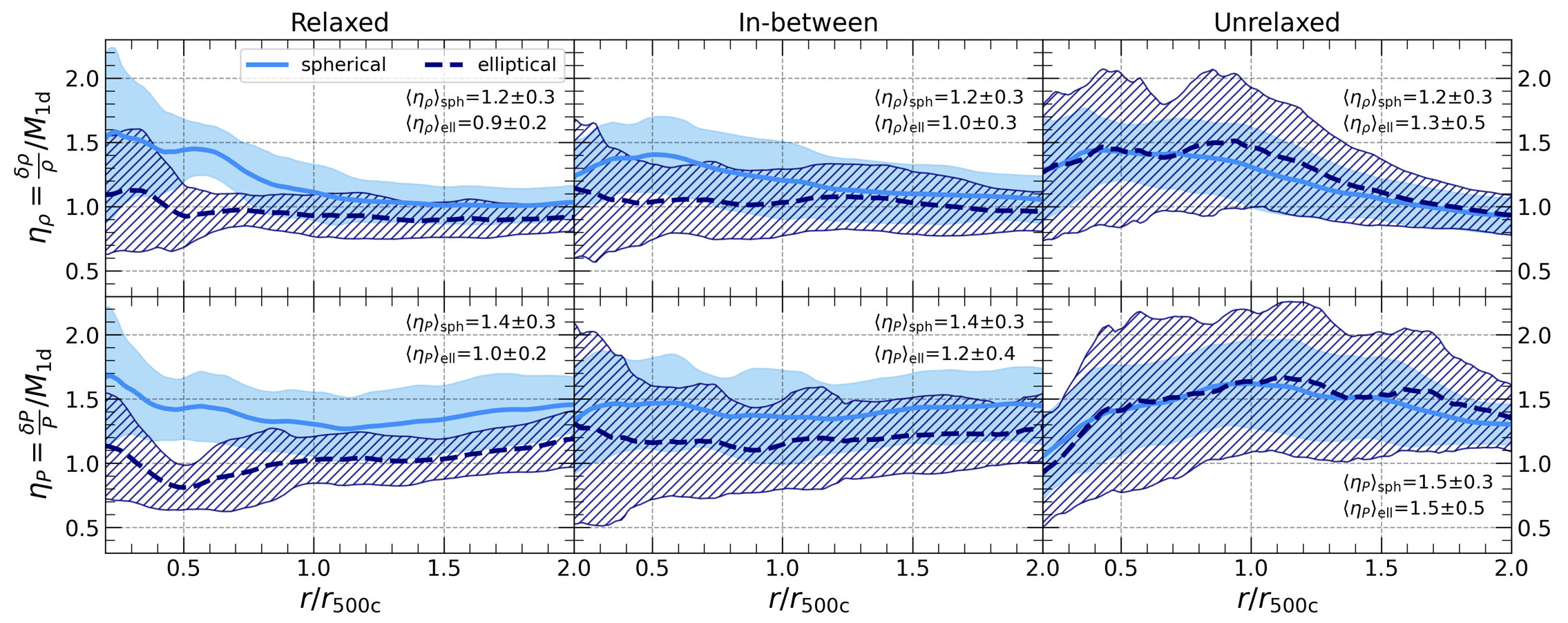}
    \caption{Radial profiles of the proportionality coefficient $\eta$ between the amplitude of density (top) or pressure (bottom) fluctuations and one-component Mach number averaged over a sample of relaxed (left), in-between (middle), and unrelaxed (right) clusters. Calculations in spherical shells are shown with light blue, solid curves/regions, while results in elliptical shells are plotted with navy, dashed/hatched curves/regions. All panels show  sample-averaged mean values and scatter. The averaged values within the entire region ($r<2r_{500c}$, accounting for the ellipticity) are written in the top-right or bottom-right corners. }
    \label{fig:eta_best}
\end{figure*}

\begin{figure*}
    \centering
    \includegraphics[width=\linewidth, trim=0 30 0 0]{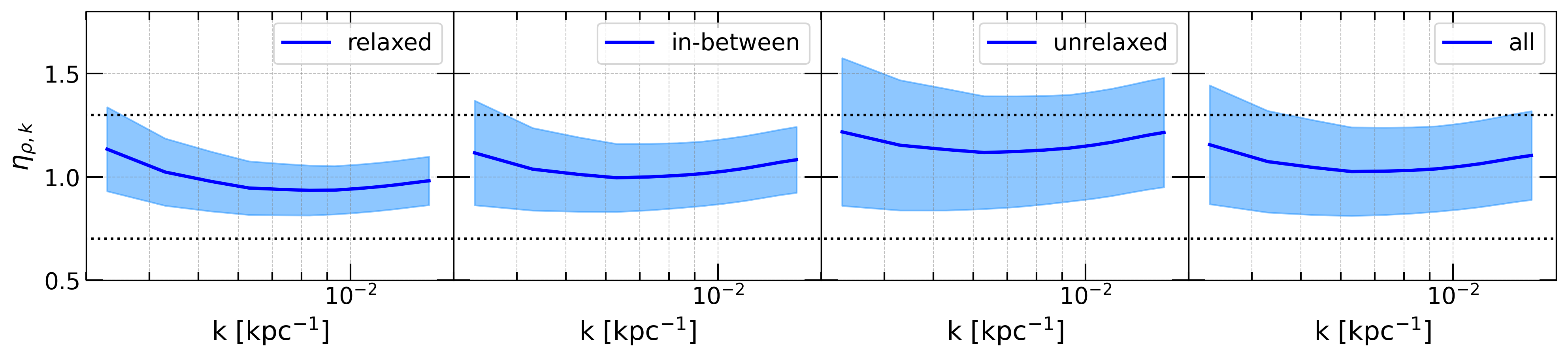}
    \caption{Proportionally coefficient between a power spectrum of density fluctuations and one-component velocity spectrum as a function of a wavenumber $k$ averaged over a sample of relaxed (left), in-between (middle left), unrelaxed (middle right), and all (right) clusters. The mean values/scatters are shown with solid curves/regions. The region between the dotted lines corresponds to earlier predictions based on a small sample of the most relaxed clusters, $\eta_{\rho,k}=1\pm 0.3$ \citep{Zhuravleva2014}. We only show wavenumbers not affected by the resolution of simulations (correspond to scales $\sim 60-300$ kpc).}
    \label{fig:eta_ps}
\end{figure*}

\begin{figure*}
    \centering
    \includegraphics[width=\linewidth, trim=0 20 0 0]{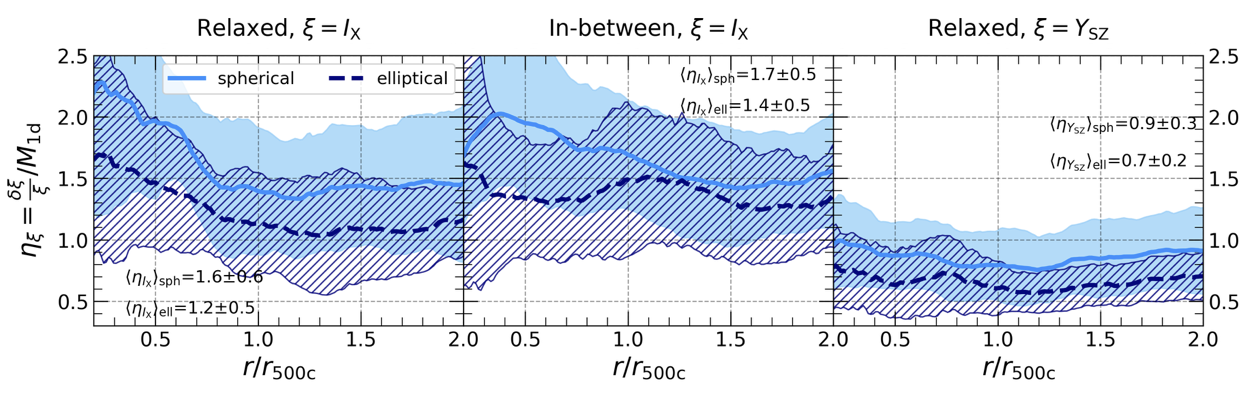}
    \caption{Proportionality coefficient $\eta$ between the amplitude of X-ray surface brightness (left, middle) or $y-$parameter (right) fluctuations  and one-component Mach number averaged over a subsample of relaxed (left, right) or in-between (middle) clusters. Notations are the same as in Fig. \ref{fig:eta_best}. }
    \label{fig:eta_2d}
\end{figure*}

Fig. \ref{fig:amplitudes} shows the widths of density, pressure, X-ray surface brightness, and $y$-parameter distributions (relative to the median value) as functions of the radius or projected radius. These widths characterize the total (i.e., integrated over all scales within a given shell or annulus) amplitude of fluctuations of the corresponding characteristics. The  calculations are done in spherical (solid, light blue) and elliptical (hatched/dashed, navy) shells or annuli and averaged over the groups of relaxed (left), in-between (middle), and unrelaxed (right) clusters. Accounting for ellipticity reduces the mean 3D amplitudes for relaxed and in-between clusters by a maximum factor of $\sim 1.9$ and $\sim 1.4$, respectively, and $\sim 1.3$ for unrelaxed ones. When measured in elliptical shells, the scatter around the mean value increases for unrelaxed clusters. This is mainly caused by the strongly asymmetric morphology of some unrelaxed clusters and large offsets between the chosen cluster center (the location of the dark matter particle with the most bounded gravitational energy) and the peak of considered gas characteristics.  The average amplitudes are below 60 (70) per cent in elliptical shells for relaxed (in-between) clusters and could be closer to 90 per cent for unrelaxed clusters. The small amplitude of density fluctuations, $\lesssim 25$ per cent, in the innermost regions is consistent with the observed amplitudes measured through X-ray surface brightness fluctuations in nearby clusters \citep[e.g.,][]{Chu12,Wal15,Zhu18} \footnote{We checked that using emissivity-weighted density, which is closer to the observational characteristic, does not affect the amplitude of density fluctuations in relaxed clusters and may only change the average amplitude in unrelaxed clusters by a small factor of $\sim 1.2$.}. The larger values of $\delta P/P$ in the outer regions are also consistent with the pressure fluctuations measured with {\it Planck} \citep{Kha16}.

While the ratio of pressure and density fluctuations amplitudes is about $\sim 1.1-1.5$ for clusters in any dynamic state, the ratio between the projected characteristics, the X-ray surface brightness, and the $y-$parameter fluctuations, is larger, a factor of $\sim 1.6-2.4$. This is not surprising given that both observational characteristics scale differently with the density. $\delta I_{\rm X}/I_{\rm X}$ and $\delta Y_{\rm SZ}/I_{\rm SZ}$ are both below $40$ per cent within $r_{500}$ for relaxed clusters and significantly larger (up to $100$ per cent in some cases) for unrelaxed systems. These conclusions are independent of the choice of projection axis. Overall, $\delta\rho/\rho$ has the smallest scatter for clusters in all dynamic states. $\delta I_{\rm X}/I_{\rm X}$, in contrast, has the largest scatter compared to fluctuations of other characteristics. This is not surprising given that $\delta I_{\rm X}/I_{\rm X}$ is most sensitive to the presence of extended structures around the high-density clumps. This is illustrated with the red dashed curves calculated using $f_{\rm cut}=2.5$ (cf. the dashed navy curves). Exclusion of more substructure around the high-density clumps affects $\delta I_{\rm X}/I_{\rm X}$ only mildly and at $r_{\rm proj}>r_{500c}$, however, suppresses the mean value of fluctuation in unrelaxed clusters by $\sim 20-30$ per cent. Note that $\delta Y_{\rm SZ}/Y_{\rm SZ}$ is not sensitive to these extended clumping structures.

One can use the amplitude of fluctuations as a proxy for the dynamical state of a cluster when classifying clusters into relaxed or unrelaxed groups. Several methods for such classification have been proposed and used in observations and simulations, including asymmetry of the X-ray surface brightness, X-ray light concentration ratio, centroid shifts, power ratios, third-order power ratio, the cross-correlation of the X-ray surface brightness and mass distribution \cite[see e.g.,][and references therein]{Buo95,Moh95,San08,We13,Nur13,Ras13,Man15,Lov17,Yua20,Del21,Cer22}. Fig. \ref{fig:amplitudes} (gray vs. navy curves/regions) clearly shows the difference between fluctuations in relaxed and dynamically active clusters. At each radius, we performed a Kolmogorov-Smirnov test to assess whether the samples of $\delta\rho/\rho$ or $\delta P/P$ values for relaxed clusters belong to a different population compared to the population of unrelaxed clusters. By generating 100 bootstrap samples, we also estimated uncertainties on the p-values (if the p-value is below 0.05 then the null hypothesis that the two samples were drawn from the same distribution is rejected). The typical p-value is smaller than $10^{-6}$ for $\delta\rho/\rho$ and $10^{-3}$ for $\delta P/P$. The lowest p-value is reached at $r=(0.5-0.7)r_{500c}$ in both elliptical and spherical shells, meaning that the pressure and density fluctuations are statistically very well separated between relaxed and unrelaxed clusters. Similar conclusions are valid when comparing relaxed and in-between clusters, however, the p-values are slightly higher. This classification method is particularly convenient for large samples of simulated galaxy clusters as the widths of density and pressure distributions can be measured robustly (e.g., not sensitive to the presence of high-density clumps) and straightforwardly. We also checked projected characteristics, finding that, overall, the p-value is $\lesssim$ 0.01 at $r<r_{500c}$ and could be higher at $r>(1-1.5)r_{500c}$. Therefore, using 3D fluctuations (density, pressure) to classify clusters based on their dynamic states is more robust compared to projected characteristics.

\subsection{Linear relation between the amplitude of gas perturbations and velocity Mach number}
\label{sec:calib}

First, we check how strong the linear correlation is between the Mach number of gas motions and all types of fluctuations in the ICM. Fig. \ref{fig:corr_all} shows a one-component Mach number vs. all the amplitudes $\delta\xi/\xi$. All characteristics are measured as a function of the elliptical radius (color-coded) within a $r\lesssim 2r_{500c}$ region (the central $0.2r_{500c}$ are excluded from the analysis as these regions are dominated by radiative physics) in each cluster. We fit all the distributions with a linear function, $M_{\rm 1d}=\alpha+\gamma(\delta\xi/\xi)$, and check the strength of each correlation through a standard Pearson's rank correlation coefficient $R_{\rm p}$. The best-fitting functions and all the parameters are plotted in blue. Relaxed clusters show a strong correlation ($R_{\rm p}\approx 0.8-0.9$) between $M_{\rm 1d}$ and density or pressure fluctuations (top two rows in Fig.~\ref{fig:corr_all}, left panels). The best-fitting line (blue curve), in this case, closely follows a one-to-one correlation (dashed gray lines) obtained earlier based on a small sample of relaxed, quasi-spherical simulated clusters \citep{Zhuravleva2014}. For in-between clusters, these correlations are weaker but still relatively strong ($R_{\rm p}\approx 0.6$). Unrelaxed clusters show a weaker correlation between $\delta\rho/\rho$ and $M_{\rm 1d}$ ($R_{\rm p}\sim 0.5$), while the correlation between $\delta P/P$ and $M_{\rm 1d}$ is still moderately strong ($R_{\rm p}\sim 0.7$). This trend for unrelaxed clusters is not surprising since gas motions are stronger with $M\gtrsim 0.5$ and, hence, the role of compressive modes is progressively increasing. ``Projected'' (i.e., potentially observable) amplitudes, $\delta I_{\rm X}/I_{\rm X}$ and $\delta Y_{\rm SZ}/Y_{\rm SZ}$, show significantly weaker correlations ($R_{\rm p}<0.5$) with $M_{\rm 1d}$ in half of the cases except for relaxed clusters ($R_{\rm p} \sim 0.5-0.7$) and, in the case of density fluctuations, for in-between ones ($R_{\rm p} \sim 0.5$). While specific values for $R_{\rm p}$ could vary slightly, our conclusions for projected characteristics qualitatively are essentially independent of the choice of projection axis and the $f_{\rm cut}$ parameter.    

Calibration of the statistical relations between the amplitudes of fluctuations and velocity at different radii, namely, $\delta\xi/\xi = \eta_{\xi}M_{\rm 1d}$, is shown in Fig. \ref{fig:eta_best}. The top panels show $\eta_{\rho}=(\delta\rho/\rho)/M_{\rm 1d}$, while the bottom ones - $\eta_{\rm P}=(\delta P/P)/M_{\rm 1d}$. All $\eta$ are averaged over subsamples of clusters at each radius (curves and regions) as well as averaged within the entire $r<2r_{500c}$ region (labels in the top- or bottom-right corners). One can see that accounting for ellipticity reduces $\eta$ in relaxed and in-between clusters, bringing it closer to the earlier-predicted value $1\pm 0.3$ \citep{Zhuravleva2014}. $\eta$ measured in elliptical shells in unrelaxed clusters has a larger scatter compared to the same calculations in radial shells, while the mean value remains almost unchanged. The dynamic state of clusters is reflected in the average scatter. Within $r<2r_{500c}$, it is the smallest for relaxed clusters ($\sim 20-30$ per cent) and could be up to $\sim 50$ per cent for unrelaxed ones. Therefore, when using $\eta$ to infer velocities from observed amplitudes of density or pressure fluctuations, it is important to take into account the ellipticity of gas distribution in relaxed and in-between clusters (affects the mean, does not change the scatter), while using spherical shells for the unrelaxed ones (does not matter for the mean $\eta$ yet the scatter is lower).

Besides the global scalings, it is interesting to check their scale-by-scale versions as they are most relevant to recently-measured velocity power spectra from the observed power spectra of density fluctuations \citep[e.g.,][]{Chu12,Wal15,Are16,Zhu18}. We checked the scaling within the central $r<0.5r_{500}$ region, where the adaptive mesh resolution of the simulations is the highest, and our resolution study showed convergence on a range of relevant wavenumbers. Following the observational procedure \citep{Are12,Chu12}, we calculated radial profiles of X-ray surface brightness for each cluster and approximated them with a $\beta-$model. Dividing the gas density by the corresponding best-fitting model, we obtained a data cube of relative density fluctuations. We then calculated the power spectra of density fluctuations and RMS velocity and took the ratio of both to get $\eta_{\rho,k}$. The averaged results for all types of clusters are shown in Fig. \ref{fig:eta_ps}. We only focus on scales from $\sim 60-300$ kpc, where the power spectra are not affected by the resolution of the simulations. Overall, the mean value of $\eta_{\rho,k}$ is consistent with global (integrated over all scales) results shown in Fig. \ref{fig:eta_best} for all types of clusters.

In observations, when reprojecting 2D (i.e., $I_{\rm X}$, $Y_{\rm SZ}$) amplitudes of fluctuations to the 3D ones (i.e., $\rho$, $P$), one relies on the assumption of many independent fluctuations of a given scale along the line of sight \citep[see Section 3 in ][]{Chu12}. Therefore, if projected amplitudes could be used to measure the velocities of gas motions, it would significantly simplify the observational procedure. Fig. \ref{fig:eta_2d} shows $\eta$ from projected amplitudes that show strong correlations with Mach number in Fig. \ref{fig:corr_all}. Interestingly, $\eta_{\rm X}$ for relaxed and in-between clusters has a scatter that is a factor of $\sim 2$ larger than for the 3D density amplitudes, while $\eta_{\rm SZ}$ for relaxed clusters has the same scatter of $\sim 20-30$ per cent as in the 3D density or pressure case. Therefore, for the relaxed clusters, all types of fluctuations, except for $\delta I_{\rm X}/I_{\rm X}$, can be used interchangeably to probe velocities of gas motions. These predictions are only slightly affected by the choice of projection axis and $f_{\rm cut}$ parameter. Depending on the choice, the mean value of $\eta$ may change by $\pm 0.1$ - still consistent with the default results within the scatter.

To summarize, the 3D amplitude of density or pressure fluctuations is a preferred proxy for velocity amplitude measurements in clusters in any dynamic state. For relaxed and in-between clusters, using elliptical shells instead of spherical ones improves the correlations between the amplitudes and Mach number. Instead, for unrelaxed clusters, using the amplitudes in spherical shells should give the tightest constraints on $M_{\rm 1d}$. The $y-$parameter fluctuations are also as robust proxies for velocities as any of the 3D amplitudes in relaxed clusters.  We also checked that using a finer resolution of cosmological simulations negligibly affects $\eta$ from 3D characteristics. The effect on mean $\eta$ from projected amplitudes within $r_{500c}$ is slightly stronger yet consistent with the fiducial case within the uncertainties.

\subsection{Improved proxies for Mach number}

\begin{figure*}
    \centering
    \includegraphics[width=\linewidth, trim=0 40 0 0]{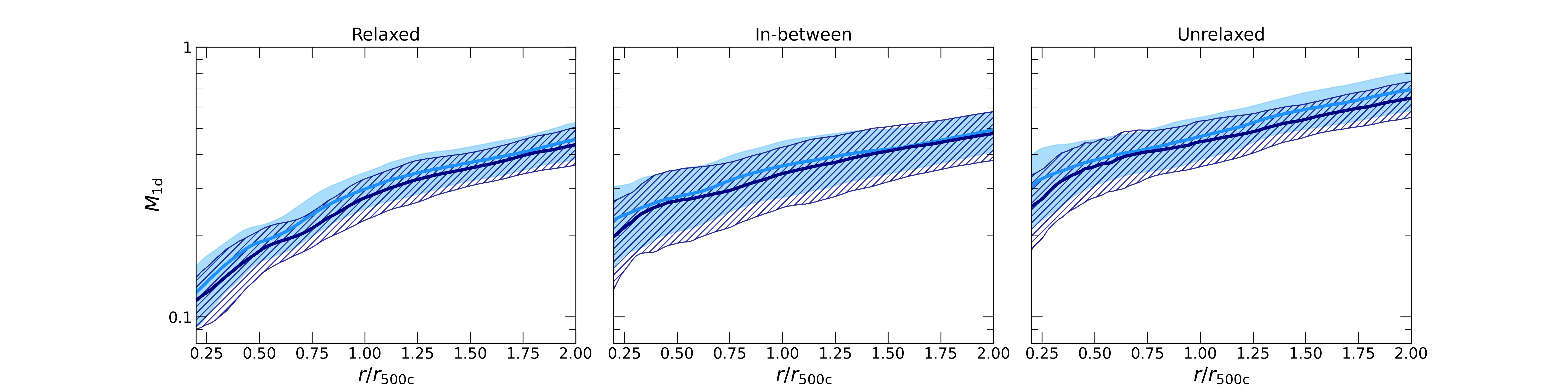}
    \caption{Radial profiles of one-component Mach number of gas motions averaged over the subsamples of relaxed, in-between, and unrelaxed clusters. Only the bulk component of the gas is considered (i.e., the high-density clumps are removed). The color coding is the same as in Fig. \ref{fig:amplitudes}. Elliptical shells are taken from density distributions.}
    \label{fig:mach_radial}
\end{figure*}

As is clear from Figs. \ref{fig:amplitudes} and \ref{fig:mach_radial}, the fluctuation amplitudes of all quantities and velocity Mach number show strong radial trends. These trends reflect a typical state of the ICM in clusters that keep growing by accretion. In other words, the gas becomes progressively more and more perturbed with the increasing radius, and the level of perturbations depends on the recent accretion history. Several interesting questions arise naturally. Are the correlations seen in Fig.~\ref{fig:corr_all} driven purely by these radial dependencies? Is the correlation between $M_{\rm 1d}$  and, e.g., density perturbations, which in stratified atmospheres can be established by processes outlined above, stronger or weaker than the typical radial trend? To what accuracy  can $M_{\rm 1d}$ be predicted using a pure radial dependence or taking into account other proxies too?

\begin{table*}
    \centering
    \begin{tabular}{c|c|c|c|c|c|c|c|c|c|c|c|c|c|c|c|c|c|c}
    \hline
      
        & $\delta\xi/\xi$ & &\multicolumn{16}{c}{$M_{\rm 1d} = \alpha + \beta (r/r_{500c})+\gamma (\delta\xi/\xi)$} \\

      Sample & $\xi=$ & $\alpha$& RMS$_{\rm i}$ & &$\alpha$ & $\beta$ & Rank & RMS & &$\alpha$ & $\gamma$ & Rank & RMS & & $\alpha$ & $\beta$ & $\gamma$ & RMS \\
       (a) & (b) & (c) &  (d) && (e) & (f) & (g)  & (h) && (i) & (j) & (k) &  (l)& & (m) & (n) & (o) & (p)\\
      
    \hline
    Relaxed & $\rho$ &0.25 & 0.09  & &  0.08  &  0.19  &  0.87  &  0.04  &  & 0.04  &  0.91  &  0.83  &  0.05  &  & 0.04  &  0.13  &  0.41  &  0.04\\
    
    & $P$ & 0.25 & 0.09  & &  0.08  &  0.19  &  0.88  &  0.04  &  & 0.08  &  0.67  &  0.86  &  0.05  &  & 0.06  &  0.11  &  0.35  &  0.04\\
    
    & $I_{\rm X}$ & 0.25 & 0.09  & &  0.08  &  0.19  &  0.87  &  0.04  &  & 0.16  &  0.32  &  0.49  &  0.08  &  & 0.06  &  0.18  &  0.09  &  0.04\\
    
    & $Y_{\rm SZ}$ & 0.25 & 0.09  & &  0.08  &  0.19  &  0.88  &  0.04  &  & 0.13  &  0.79  &  0.66  &  0.07  &  & 0.06  &  0.16  &  0.26  &  0.04\\
    
    \hline
   In-betw. & $\rho$ &0.33 & 0.11  & &  0.19  &  0.15  &  0.59  &  0.09  &  & 0.15  &  0.55  &  0.59  &  0.09  &  & 0.11  &  0.1  &  0.36  &  0.08 \\

    & $P$& 0.33 & 0.10  & &  0.19  &  0.15  &  0.62  &  0.08  &  & 0.19  &  0.36  &  0.58  &  0.08  &  & 0.13  &  0.11  &  0.24  &  0.07\\

    & $I_{\rm X}$ &0.33 & 0.11  & &  0.19  &  0.15  &  0.59  &  0.09  &  & 0.21  &  0.26  &  0.52  &  0.09  &  & 0.14  &  0.12  &  0.17  &  0.08\\
    
    & $Y_{\rm SZ}$ &0.33 & 0.10  & &  0.19  &  0.15  &  0.62  &  0.08  &  & 0.25  &  0.34  &  0.38  &  0.10  &  & 0.16  &  0.14  &  0.17  &  0.08\\

    \hline
    Unrel. & $\rho$ &0.43 & 0.11  & &  0.26  &  0.19  &  0.70  &  0.08  &  & 0.29  &  0.25  &  0.48  &  0.10  &  & 0.18  &  0.17  &  0.16  &  0.07\\
    
    & $P$ &0.43 & 0.12  & &  0.25  &  0.20  &  0.71  &  0.08  &  & 0.28  &  0.23  &  0.67  &  0.09  &  & 0.21  &  0.14  &  0.14  &  0.07\\
    
    & $I_{\rm X}$ &0.43 & 0.11  & &  0.26  &  0.19  &  0.70  &  0.08  &  & 0.38  &  0.07  &  0.27  &  0.11  &  & 0.23  &  0.18  &  0.04  &  0.08\\
    
    & $Y_{\rm SZ}$ &0.43 & 0.12  & &  0.25  &  0.20  &  0.71  &  0.08  &  & 0.35  &  0.19  &  0.42  &  0.11  &  & 0.22  &  0.18  &  0.10  &  0.08 \\
   
    \hline
    \hline
           & $\delta\xi/\xi$ & &\multicolumn{16}{c}{$M_{\rm 1d} = \alpha   (r/r_{500c})^{\beta}(\delta\xi/\xi)^{\gamma}$} \\

      Sample & $\xi=$ & $\alpha$& RMS$_{\rm i}$ & &$\alpha$ & $\beta$ & Rank & RMS & &$\alpha$ & $\gamma$ & Rank & RMS & & $\alpha$ & $\beta$ & $\gamma$ & RMS \\
       (a) & (b) & (c) &  (d) && (e) & (f) & (g)  & (h) && (i) & (j) & (k) &  (l)& & (m) & (n) & (o) & (p)\\
      
    \hline
    Relaxed & $\rho$ & 0.25 & 0.09  & &  0.27  &  0.63  &  0.89  &  0.04  &  & 0.83  &  0.83  &  0.81  &  0.05  &  & 0.4  &  0.47  &  0.29  &  0.04\\

    & $P$ & 0.25 & 0.09  & &  0.27  &  0.63  &  0.89  &  0.04  &  & 0.65  &  0.68  &  0.85  &  0.04  &  & 0.4  &  0.42  &  0.29  &  0.04\\
    
    & $I_{\rm X}$ &  0.25 & 0.09  & &  0.27  &  0.63  &  0.89  &  0.04  &  & 0.41  &  0.42  &  0.48  &  0.08  &  & 0.3  &  0.6  &  0.1  &  0.04\\
    
    & $Y_{\rm SZ}$ & 0.25 & 0.09  & &  0.27  &  0.63  &  0.89  &  0.04  &  & 0.63  &  0.50  &  0.64  &  0.07  &  & 0.33  &  0.56  &  0.11  &  0.04\\
    \hline
   In-betw. & $\rho$ &0.33 & 0.11  & &  0.34  &  0.37  &  0.65  &  0.09  &  & 0.63  &  0.60  &  0.67  &  0.08  &  & 0.51  &  0.22  &  0.39  &  0.08\\
   
    & $P$& 0.33 & 0.10  & &  0.34  &  0.38  &  0.67  &  0.08  &  & 0.50  &  0.45  &  0.65  &  0.08  &  & 0.45  &  0.26  &  0.28  &  0.07\\
    
    & $I_{\rm X}$ &0.33 & 0.11  & &  0.34  &  0.37  &  0.65  &  0.09  &  & 0.43  &  0.35  &  0.52  &  0.09  &  & 0.4  &  0.30  &  0.19  &  0.08\\
    
    & $Y_{\rm SZ}$ &0.33 & 0.10  & &  0.34  &  0.38  &  0.67  &  0.08  &  & 0.49  &  0.28  &  0.46  &  0.09  &  & 0.42  &  0.33  &  0.14  &  0.08\\
    \hline
    Unrel. & $\rho$ &0.43 & 0.11  & &  0.45  &  0.37  &  0.70  &  0.08  &  & 0.55  &  0.42  &  0.58  &  0.10  &  & 0.52  &  0.30  &  0.25  &  0.08\\
    
    & $P$ &0.43 & 0.12  & &  0.45  &  0.38  &  0.70  &  0.08  &  & 0.52  &  0.41  &  0.74  &  0.08  &  & 0.50  &  0.21  &  0.27  &  0.07\\
    
    & $I_{\rm X}$ &0.43 & 0.11  & &  0.45  &  0.37  &  0.70  &  0.08  &  & 0.44  &  0.18  &  0.36  &  0.11  &  & 0.46  &  0.35  &  0.09  &  0.08\\
    
    & $Y_{\rm SZ}$ & 0.43 & 0.12  & &  0.45  &  0.38  &  0.70  &  0.08  &  & 0.52  &  0.21  &  0.48  &  0.10  &  & 0.49  &  0.33  &  0.09  &  0.08\\
    \hline
    \end{tabular}
    \caption{Summary of correlations between the distributions of Mach number and the amplitude of fluctuations, and Mach and radius. (a): subsample used for calculations; (b): fluctuating characteristic; (c-d): the mean value of Mach number and RMS of Mach number around the mean value calculated from the initial $M_{\rm 1d} - \delta\xi/\xi$ correlation (see Fig. \ref{fig:corr_all}); (e) - (h): the best-fitting parameters of the correlation between $M_{\rm 1d}$ and radius $r/r_{500c}$, Pearson's rank, and the RMS of Mach number after subtracting this best-fitting model; (i)-(l): the same but for the correlation between $M_{\rm 1d}$ and $\delta\xi/\xi$, see also Fig. \ref{fig:corr_all}; (m) - (p): the best fitting parameters of the correlation between $M_{\rm 1d}$, $r/r_{500c}$ and $\delta\xi/\xi$, and the RMS of Mach number once this best-fitting model is subtracted from the correlation. Note that for each subsample, the values (e-h) are almost the same for all types of fluctuations because the ellipticities of these characteristics are very similar. The top half of the table shows results for linear relations, while the bottom half summarizes correlations between log-10s of the considered characteristics (i.e., exploring a power law correlation).}
    \label{tab:m1d_correlations}
\end{table*}

To answer these questions, we checked all the correlations, summarizing the results in Table \ref{tab:m1d_correlations}. For all three subsamples, we calculated the mean and RMS of the Mach number relative to the mean value from the initial $M_{\rm 1d} - \delta\xi/\xi$ correlations [columns (c-d)]. Not surprisingly, the mean $M_{\rm 1d}$ is increasing with the dynamic state of a cluster, from $\sim 0.25$ for relaxed ones and reaching $\sim 0.43$, on average, for unrelaxed ones. It is remarkable, though, that the scatter around the mean [column (d)] is relatively small, $\sim 0.1$, and is essentially the same for all groups.

Columns (e)-(h) show results for the $M_{\rm 1d} - r/r_{500c}$ correlation, namely, the best fitting intercepts and slopes, Pearson's rank, and the RMS of Mach number after this best-fitting model is removed from the correlation. Relaxed clusters show a strong correlation of $M_{\rm 1d}$ with radius, $R_{\rm p} \sim 0.9$. For perturbed clusters, the correlation is weaker but still within the strong range, $R_{\rm p} \sim 0.6-0.7$. Once the best-fitting models are subtracted from the correlations, the Mach number RMS reduces by a factor of $\sim 2/1.3/1.5$ for relaxed/in-between/unrelaxed clusters, respectively.

As for the $M_{\rm 1d} - \delta\xi/\xi$ [columns (i)-(l), the distributions are shown in Fig. \ref{fig:corr_all}], one can see that the correlations with 3D amplitudes (density or pressure) are essentially as strong as with radius in relaxed clusters, $R_{\rm p}\sim 0.8-0.9$. This means that both correlations could be used interchangeably. When fitting $M_{\rm 1d}$ with two parameters ($r$ and $\delta\xi/\xi$), it does not reduce the RMS further. ``Projected'' amplitudes correlate weaker with $M_{\rm 1d}$, especially $\delta I_{\rm X}/I_{\rm X}$ (as we saw in Fig. \ref{fig:corr_all}), and only reduce the initial RMS by 0.01-0.02. For in-between clusters, the correlation with 3D amplitudes is almost as strong as with $r/r_{500c}$, and combining all three characteristics reduced the RMS the most, by a factor of $\sim 1.6$. In contrast to the relaxed group, the correlation with $\delta I_{\rm X}/I_{\rm X}$ is significantly stronger than with $\delta Y_{\rm SZ}/Y_{\rm SZ}$. Unrelaxed clusters show the strongest correlation with the amplitude of pressure fluctuations, comparable with the correlation strength with the radius. When radial correlations are combined with the 3D amplitudes, the initial RMS is reduced the most, by a factor of $\sim 1.6-1.7$.

A good performance of the simple (linear) radial model is encouraging, given its simplicity. Verifying these predictions with X-ray and sub-mm observations via measuring X-ray surface brightness and/or projected pressure fluctuations and converting them to 3D proxies would be very interesting. Since the $M_{\rm 1d} - r$ correlation is driven by the history of structure formation while $M_{\rm 1d} - \delta\xi/\xi$ by intrinsic fluid properties within the stratified ICM, measuring Mach numbers through both proxies and comparing them with direct velocity measurements could provide important constraints for cosmological models of galaxy clusters.

Now that we found the best-fitting correlations of $M_{\rm 1d}$ and $\delta\xi/\xi$ with the radius, it is interesting to check a residual $M_{\rm 1d} - \delta\xi/\delta\xi$ correlation, i.e., the correlation between these characteristics after the corresponding radial trends are subtracted. This is shown in Fig. \ref{fig:corr_nor} in Appendix A. One can see that the correlations with $\delta\rho/\rho$ and $\delta P/P$ remain moderately strong for relaxed clusters ($R_{\rm p}\sim 0.5-0.6$), while become weaker for in-between and unrelaxed clusters (yet, at a moderate level, $R_{\rm p}\sim 0.4 - 0.5$). This exercise confirms that, at least for weakly perturbed clusters, the amplitude of density/pressure perturbations bears additional information on the gas velocities that are not captured by the pure radial trends.

As is seen from Table \ref{tab:m1d_correlations}, the remaining scatter of the Mach number around the model that uses two-parameters fits increases from $\sim 0.04$ for the relaxed clusters to $\sim 0.08$ for the in-between and unrelaxed sub-samples. If the sample-averaged value is of interest, this factor of 2 increase in the RMS could be compensated by a factor of 4 larger samples, so that the factor $1/\sqrt{N}$, where $N$ is the number of objects in the sample, compensates for the increased RMS. Since the relaxed sample makes up about a quarter of all clusters, there is a certain value in dealing with the entire sample. However, in practice, working with a cleaner and smaller sample has many other benefits and appears as a viable option.

Besides the linear correlations, it is interesting to check a linear scaling between the logarithms of the characteristics or, equivalently, a power-law scaling between the initial quantities. Red lines and parameters in Fig. \ref{fig:corr_all} show the best-fitting results between the log10 characteristics. The second half of Table \ref{tab:m1d_correlations} also summarizes all considered cases. For the correlations with radius and $\delta\xi/\xi$, both linear and power-law scalings provide consistent results: the same cases of strong correlations, RMS is reduced by similar factors. The amplitude $\delta\rho/\rho \sim M_{\rm 1d}^{1.2}$ for relaxed clusters, i.e., close to the confirmed linear scaling, while for unrelaxed clusters, the scaling is closer to $\sim M_{\rm 1d}^2$ (namely, $\delta\rho/\rho$ and $\delta P/P$ are $\sim M_{\rm 1d}^{2.4}$). Idealized hydrodynamic simulations of turbulence in stratified cluster atmospheres in a static gravitational potential with pure solenoidal driving \citep{Gas14,Moh20} already showed that such scaling is expected at $M_{\rm 1d} > 0.3$. It is interesting that despite all the complexity of structure formation, clusters in cosmological simulations confirm this result.

\section{Discussion}
\label{sec:discussion}
\subsection{Proxies for hydrostatic mass bias}

\begin{figure*}
    \centering
    \includegraphics[width=\linewidth, trim=0 50 0 0 ]{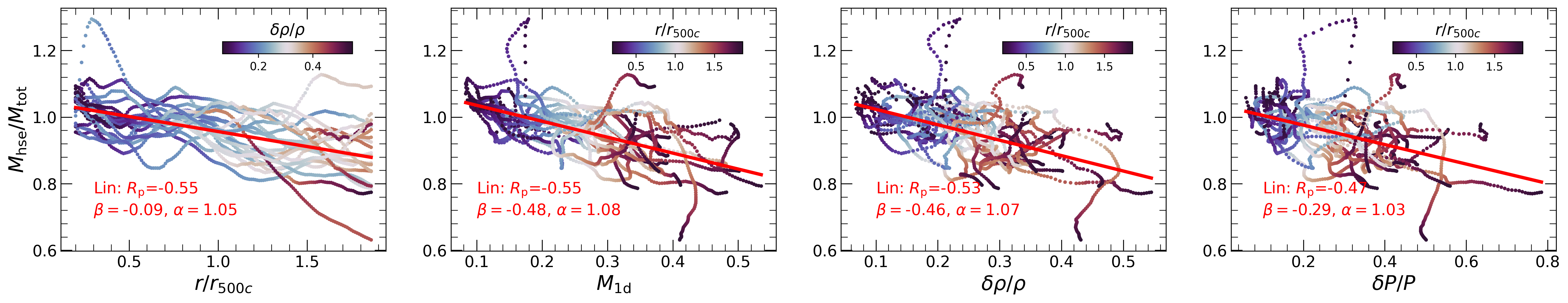}
    \caption{Correlations between the mass bias, $M_{\rm hse}/M_{\rm tot}$, and radius (left), Mach number of gas motions (middle left), the amplitude of density (middle right) or pressure (right) fluctuations in a sample of relaxed clusters. The amplitudes and Mach numbers are calculated in elliptical shells. The color indicates the amplitude of density fluctuations in the left and radius in all other panels. Each track of points corresponds to an individual cluster in the sample. Red lines are the best fitting linear functions, $M_{\rm hse}/M_{\rm tot}=\alpha+\beta X$, where $X$ is the corresponding characteristic on the X-axis. The best-fitting $\alpha, \beta$ and Pearson's ranks, $R_{\rm p}$, are shown in legends.}
    \label{fig:mbias}
\end{figure*}

Measuring velocities of gas motions in the ICM is important for precise cluster mass measurements through their X-ray or SZ observations under the assumption of hydrostatic equilibrium between the thermal pressure of the gas and gravity. Gas motions may provide significant non-thermal pressure support, up to 10-20 per cent at $r_{500}$ \citep[e.g.,][]{Lau09,Nelson2012,Nelson2014,Nelson2014b}, even in relaxed clusters, and should be taken into account for the mass measurements suitable for high-precision cosmology \citep[][for a recent review]{Pratt2019}. 

In the context of this work, it is interesting to examine various approximate proxies for mass bias, namely, the Mach number of gas motions (that will be observed soon, e.g., with {\it XRISM}) and various amplitudes of fluctuations (can be probed with current X-ray and SZ imaging data), and compare them with predictions from the radial trends of mass bias in cosmological simulations. We define the mass bias as $M_{\rm bias}=M_{\rm hse}/M_{\rm tot}$, where $M_{\rm hse}$ is the mass derived from the equation of hydrostatic equilibrium, namely, $M_{\rm hse}(<r)=-\frac{r^2}{G\rho(r)}\frac{dP_{\rm th}(r)}{dr}$, where $\rho(r)$ and $P_{\rm th}(r)$ are the radial profiles of gas density and thermal pressure, respectively, and $M_{\rm tot}$ is the total mass enclosed within a sphere of radius $r$. Fig. \ref{fig:mbias} shows the correlations of the mass bias with radius, Mach number, and the amplitudes of density and pressure fluctuations. As before, we measured these characteristics in each cluster at each radius. Only relaxed clusters are considered, and calculations are done within $\approx 2r_{500c}$, excluding the central $0.2r_{500c}$. In all four cases, the correlations are moderately strong, with Pearson's rank $R_{\rm p} \sim 0.5$. Not surprisingly, the best correlation of $M_{\rm bias}$ is found with the Mach number, although a pure radial trend performs almost equally well. Among all the considered amplitudes of fluctuations, $\delta\rho/\rho$ shows the strongest correlation with $M_{\rm bias}$, $R_{\rm p}\approx 0.53$, while ``projected'' amplitudes correlate very weakly (not shown in the Figure). We fit all correlations with a linear function, $M_{\rm bias}=\alpha+\beta X$, where $X$ is one of the considered characteristics on the X-axes. The best-fitting parameters and functions are shown in red. One can see that the global trends of $M_{\rm bias}$ could be easily removed using any of the four characteristics. As for the scatter (RMS of $M_{\rm bias}$ around the mean value), it is reduced from the initial value $\sim 8$ per cent to $\sim 6$ per cent when removing the correlation with the Mach number. For the radius, $\delta\rho/\rho$ and $\delta P/P$, the residual RMS in $M_{\rm bias}$ amounts to $\sim 7$ per cent.  

This result shows that, on average, it is possible to correct hydrostatic masses for mass bias for a sample of relaxed galaxy clusters using any of the discussed observables (Mach number, $\delta\rho/\rho$, $\delta P/P$). The corrections based on 3D amplitudes work almost as well as the $M_{\rm 1d}$ correction, i.e., the remaining scatter after any of these corrections is about the same $\sim 6-7$ per cent. We also checked if combining multiple proxies for the mass bias may further reduce the scatter (i.e., performing two-parameter fits to $M_{\rm bias}$), however, the scatter always remained at the same level. The origin of this residual scatter requires further studies and is beyond the scope of this study. Note that measuring radial information (radial profile of velocity or the amplitude of fluctuations) is crucial since local (i.e., within smaller radial regions) correlations are significantly weaker. The success of the radial trends of $M_{\rm bias}$ (Fig. \ref{fig:mbias}, left) is encouraging. Indeed, if confirmed and calibrated with the X-ray and SZ imaging and XRISM measurements for $M_{\rm 1d}$, it will represent the easiest way of making the first-order hydrostatic bias corrections to derived masses.

\subsection{Caveats and future studies}

Our results are based on a sample of clusters simulated with minimal (non-radiative) physics. Any additional physics, e.g., gas cooling, feedback process, MHD, may affect gas clumpiness, the amplitude of fluctuations, and considered correlations. Moreover, various prescriptions for these additional physics in cosmological simulations may also have an effect. Therefore, it is important to extend the analysis to other cosmological simulations in the future. 

To approximately estimate the robustness of our conclusions, we analyzed a subsample of our clusters that were modeled in a cosmological context, including additional physics (cooling, star formation, SN and AGN feedback, UV background). While the mean amplitudes of fluctuation at specific radii may change maximum by a factor of $\sim 1.5-2$, they are consistent with the NR results within the scatter. Projected amplitudes show a slightly increased scatter with the additional physics included, while the mean values remain consistent. The main conclusions from the correlations with $M_{\rm 1d}$ are qualitatively the same for both types of simulations, and the proportionality coefficients $\eta$ agree well within the scatters. Overall, the implemented additional physics in the \textsc{Omega500} simulations does not seem to affect our main conclusions.

It is important to mention that any comparisons of these types of studies have to take into consideration that adopted procedures for velocity filtering (e.g., decomposition of the velocity field into coherent and turbulent flows) may lead to substantial differences in results. The results also depend quite strongly on the choice of considered regions and ellipticity treatment in the calculations.

Finally, it is important to mention that, to some extent, our results rely on a somewhat subjective, visual classification of clusters based on their dynamical state (relaxed, in-between, unrelaxed). Other classifications of the same sample have been used in previous works \citep[see e.g.,][]{Shi16,Chen2019}. Comparison with the cluster classification based on the mass ratio of mergers within the last 4.5 Gyr by \citet{Chen2019} shows that while the boundary between unrelaxed and in-between clusters could be blurred, the selection of relaxed clusters is quite robust. Indeed, about 84 per cent of relaxed clusters, identified based on our visual classification, were classified as smooth mergers (equivalent to relaxed) from the mass accretion histories. Overall, the classification of more than 60 per cent of clusters in our sample is weakly dependent on the chosen classification method. This means that our main results on relaxed vs. perturbed clusters should be stable to the choice of the classification method.

\section{Conclusions}
\label{sec:conclusions}

In this study, we explored gas density and pressure fluctuations in the ICM as well as their directly observable characteristics, namely fluctuations of X-ray surface brightness and SZ $y-$parameter, using a sample of 78 galaxy clusters from cosmological simulations \textsc{Omega500}. We examined scaling relations between the amplitudes of these fluctuations and velocities of gas motions in clusters in different dynamic states (relaxed, in-between, and unrelaxed), accounting for radial variations of these characteristics and halo ellipticity. We further calibrated a commonly-used in observations statistical relation between the amplitude of fluctuations and velocity Mach number in the central and outer cluster regions. We studied whether the amplitude of density fluctuations provides a much more accurate proxy for velocity amplitudes compared to a mean radial dependence of amplitudes with radius. Finally, we discussed whether these characteristics could be used as reliable proxies for hydrostatic mass bias and the expected level of remaining scatter in the bias. Our main findings are summarized below:  
\begin{itemize}
    \item{With a large sample of simulated galaxy clusters, we confirm that the averaged over the subsamples of relaxed, in-between, and unrelaxed clusters amplitudes of fluctuations increase with the distance from the cluster center. The sample-averaged amplitude of density or pressure fluctuations in relaxed clusters is below 60 per cent, while it could be up to 90 per cent in unrelaxed ones. ``Projected'' amplitudes, $\delta I_{\rm X}/I_{\rm X}$ and $\delta Y_{\rm SZ}/Y_{\rm SZ}$ integrated over all spatial scales, vary between $\sim 0.2-0.5$ and $\sim 0.1-0.3$ at $r=(0.2-2)r_{500c}$ in relaxed objects and between $\sim 0.6-1$ and $0.25-0.7$ in unrelaxed ones, respectively. Accounting for halo ellipticity is important, especially for relaxed clusters within the central $\sim r_{500c}$ regions, as it may reduce the average amplitude by up to a factor of $\sim 2$ at some radii. For unrelaxed clusters, halo ellipticity almost does not affect the mean amplitude, while increasing the scatter (Fig. \ref{fig:amplitudes}). These predictions could be checked with current X-ray and SZ observations using cluster imaging data.}
    \item{Among the considered amplitudes of fluctuations, the sample-averaged mean $\delta\rho/\rho$ at a given $r$ has the lowest scatter in clusters in all dynamic states. The scatter in $\delta I_{\rm X}/I_{\rm X}$ appears the largest, however, it is most prone to the method and parameters used to remove high-density clumps from the data.}
    \item {One can use the amplitude of density or pressure fluctuations for classifying galaxy clusters based on their dynamic state. Relaxed, in-between, and unrelaxed clusters are statistically separated especially well at $r\sim (0.5-0.7)r_{500c}$ in both spherical and elliptical shells. This classification can be performed for both simulated and observed galaxy clusters.}
    \item{There is a strong linear, almost one-to-one,  correlation between the distributions of $\delta\rho/\rho$ (or $\delta P/P$) and the Mach number of gas motions in relaxed galaxy clusters within the $2r_{500c}$ region. For perturbed clusters, the correlation is weaker but still within the moderately strong regime. This is encouraging as one can use the measured amplitudes as robust proxies for velocities of gas motions in galaxy clusters in various dynamic states. The average proportionality coefficient between the amplitude of density (pressure) fluctuations and Mach number is $0.9\pm 0.2$ ($1\pm0.2$), $1\pm0.3$ ($1.2\pm0.4$), and $1.3\pm0.5$ ($1.5\pm0.5$) for relaxed, in-between and unrelaxed clusters, respectively. When using this method, it is important to account for the ellipticity of gas distribution, especially for relaxed and mildly perturbed clusters.}
    \item{Linear correlations between ``projected'' amplitudes and Mach number are significantly weaker (Pearson ranks $<0.5$) except for $\delta I_{\rm X}/I_{\rm X}$ for in-between clusters and $\delta Y_{\rm SZ}/Y_{\rm SZ}$ for relaxed ones. While these ``projected'' amplitudes could be directly measured from the X-ray and SZ images, it is important to convert them to $\delta \rho/\rho$ or $\delta P/P$ for the most robust constraints on $M_{\rm 1d}$. This conversion is non-trivial and requires further studies, however, the idea could be similar to the one proposed for scale-dependent amplitudes and applied to observed power spectra of X-ray surface brightness fluctuations \citep[see][for details]{Chu12,Zhu15}.}
    \item{The power-law scalings between $M_{\rm 1d}$, radius, and the amplitude of fluctuations are almost as strong as the linear correlations. They also lead to similar-level scatter in residual Machs as the linear models once removed from the initial data. For relaxed clusters, a power-law scaling between $\delta\rho/\rho$ and $M_{\rm 1d}$ is close to the linear one, while unrelaxed clusters show the best-fitting scaling closer to $\delta\rho/\rho $ (the same for $\delta P/ P$) $\propto M_{\rm 1d}^2$, indicating an increased role of compressive modes and/or strong variations of the perturbed gravitational potential in unrelaxed systems. }
    \item{The hydrostatic mass bias calculated in relaxed galaxy clusters at each radius within the $r\sim 2r_{500c}$ region correlates moderately strongly with the radius, Mach number, and the amplitudes of density or pressure fluctuations, with the Pearson rank $\sim 0.5$. One can use the provided best-fitting scalings to correct for the average mass bias in observations. The remaining  scatter is expected $\sim 6$ per cent when using the correlation with $M_{\rm 1d}$ and $\sim 7$ per cent if any of the other three correlations is used. If the radial trend of the mass bias predicted in simulations is confirmed with observations, it would provide the easiest way of making the first-order hydrostatic mass bias correction to the observed cluster masses. }
\end{itemize}
Measuring Mach numbers through various proxies could provide important constraints for cosmological models since the $M_{\rm 1d} - r$ correlation is driven by the history of structure formation while $M_{\rm 1d} - \delta\xi/\xi$ by intrinsic fluid properties within the ICM. Current X-ray and SZ observations of galaxy clusters can verify our predictions on gas fluctuations, while soon-launched {\it XRISM} will provide velocities necessary for testing the remaining predictions.

\section*{Acknowledgements}
Part of the analyses presented in this paper were carried out using the Midway computing cluster provided by the University of Chicago Research Computing Center. IZ is partially supported by a Clare Boothe Luce Professorship from the Henry Luce Foundation.

\section*{Data Availability}

The data underlying this article will be shared on reasonable request to the corresponding author.




\bibliographystyle{mnras}
\bibliography{current} 

\begin{thebibliography}{}
\makeatletter
\relax
\def\mn@urlcharsother{\let\do\@makeother \do\$\do\&\do\#\do\^\do\_\do\%\do\~}
\def\mn@doi{\begingroup\mn@urlcharsother \@ifnextchar [ {\mn@doi@}
  {\mn@doi@[]}}
\def\mn@doi@[#1]#2{\def\@tempa{#1}\ifx\@tempa\@empty \href
  {http://dx.doi.org/#2} {doi:#2}\else \href {http://dx.doi.org/#2} {#1}\fi
  \endgroup}
\def\mn@eprint#1#2{\mn@eprint@#1:#2::\@nil}
\def\mn@eprint@arXiv#1{\href {http://arxiv.org/abs/#1} {{\tt arXiv:#1}}}
\def\mn@eprint@dblp#1{\href {http://dblp.uni-trier.de/rec/bibtex/#1.xml}
  {dblp:#1}}
\def\mn@eprint@#1:#2:#3:#4\@nil{\def\@tempa {#1}\def\@tempb {#2}\def\@tempc
  {#3}\ifx \@tempc \@empty \let \@tempc \@tempb \let \@tempb \@tempa \fi \ifx
  \@tempb \@empty \def\@tempb {arXiv}\fi \@ifundefined
  {mn@eprint@\@tempb}{\@tempb:\@tempc}{\expandafter \expandafter \csname
  mn@eprint@\@tempb\endcsname \expandafter{\@tempc}}}

\bibitem[\protect\citeauthoryear{{Angelinelli}, {Vazza}, {Giocoli}, {Ettori},
  {Jones}, {Brunetti}, {Br{\"u}ggen}  \& {Eckert}}{{Angelinelli}
  et~al.}{2020}]{Angelinelli20}
{Angelinelli} M.,  {Vazza} F.,  {Giocoli} C.,  {Ettori} S.,  {Jones} T.~W.,
  {Brunetti} G.,  {Br{\"u}ggen} M.,   {Eckert} D.,  2020, \mn@doi [\mnras]
  {10.1093/mnras/staa975}, \href
  {https://ui.adsabs.harvard.edu/abs/2020MNRAS.495..864A} {495, 864}

\bibitem[\protect\citeauthoryear{{Ar{\'e}valo}, {Churazov}, {Zhuravleva},
  {Hern{\'a}ndez-Monteagudo}  \& {Revnivtsev}}{{Ar{\'e}valo}
  et~al.}{2012}]{Are12}
{Ar{\'e}valo} P.,  {Churazov} E.,  {Zhuravleva} I.,  {Hern{\'a}ndez-Monteagudo}
  C.,   {Revnivtsev} M.,  2012, \mn@doi [\mnras]
  {10.1111/j.1365-2966.2012.21789.x}, \href
  {https://ui.adsabs.harvard.edu/abs/2012MNRAS.426.1793A} {426, 1793}

\bibitem[\protect\citeauthoryear{{Ar{\'e}valo}, {Churazov}, {Zhuravleva},
  {Forman}  \& {Jones}}{{Ar{\'e}valo} et~al.}{2016}]{Are16}
{Ar{\'e}valo} P.,  {Churazov} E.,  {Zhuravleva} I.,  {Forman} W.~R.,   {Jones}
  C.,  2016, \mn@doi [\apj] {10.3847/0004-637X/818/1/14}, \href
  {https://ui.adsabs.harvard.edu/abs/2016ApJ...818...14A} {818, 14}

\bibitem[\protect\citeauthoryear{{Avestruz}, {Lau}, {Nagai}  \&
  {Vikhlinin}}{{Avestruz} et~al.}{2014}]{Avestruz2014}
{Avestruz} C.,  {Lau} E.~T.,  {Nagai} D.,   {Vikhlinin} A.,  2014, \mn@doi
  [\apj] {10.1088/0004-637X/791/2/117}, \href
  {https://ui.adsabs.harvard.edu/abs/2014ApJ...791..117A} {791, 117}

\bibitem[\protect\citeauthoryear{{Barnes}, {Vogelsberger}, {Pearce}, {Pop},
  {Kannan}, {Cao}, {Kay}  \& {Hernquist}}{{Barnes} et~al.}{2021}]{Barnes21}
{Barnes} D.~J.,  {Vogelsberger} M.,  {Pearce} F.~A.,  {Pop} A.-R.,  {Kannan}
  R.,  {Cao} K.,  {Kay} S.~T.,   {Hernquist} L.,  2021, \mn@doi [\mnras]
  {10.1093/mnras/stab1276}, \href
  {https://ui.adsabs.harvard.edu/abs/2021MNRAS.506.2533B} {506, 2533}

\bibitem[\protect\citeauthoryear{{Battaglia}, {Bond}, {Pfrommer}  \&
  {Sievers}}{{Battaglia} et~al.}{2012}]{Bat12}
{Battaglia} N.,  {Bond} J.~R.,  {Pfrommer} C.,   {Sievers} J.~L.,  2012,
  \mn@doi [\apj] {10.1088/0004-637X/758/2/74}, \href
  {https://ui.adsabs.harvard.edu/abs/2012ApJ...758...74B} {758, 74}

\bibitem[\protect\citeauthoryear{{Biffi} et~al.,}{{Biffi} et~al.}{2016}]{Bif16}
{Biffi} V.,  et~al., 2016, \mn@doi [\apj] {10.3847/0004-637X/827/2/112}, \href
  {https://ui.adsabs.harvard.edu/abs/2016ApJ...827..112B} {827, 112}

\bibitem[\protect\citeauthoryear{{Bonafede} et~al.,}{{Bonafede}
  et~al.}{2018}]{Bon18}
{Bonafede} A.,  et~al., 2018, \mn@doi [\mnras] {10.1093/mnras/sty1121}, \href
  {https://ui.adsabs.harvard.edu/abs/2018MNRAS.478.2927B} {478, 2927}

\bibitem[\protect\citeauthoryear{{Brunetti} \& {Lazarian}}{{Brunetti} \&
  {Lazarian}}{2007}]{Bru07}
{Brunetti} G.,  {Lazarian} A.,  2007, \mn@doi [\mnras]
  {10.1111/j.1365-2966.2007.11771.x}, \href
  {https://ui.adsabs.harvard.edu/abs/2007MNRAS.378..245B} {378, 245}

\bibitem[\protect\citeauthoryear{{Buote} \& {Tsai}}{{Buote} \&
  {Tsai}}{1995}]{Buo95}
{Buote} D.~A.,  {Tsai} J.~C.,  1995, \mn@doi [\apj] {10.1086/176326}, \href
  {https://ui.adsabs.harvard.edu/abs/1995ApJ...452..522B} {452, 522}

\bibitem[\protect\citeauthoryear{{Cerini}, {Cappelluti}  \&
  {Natarajan}}{{Cerini} et~al.}{2022}]{Cer22}
{Cerini} G.,  {Cappelluti} N.,   {Natarajan} P.,  2022, arXiv e-prints, \href
  {https://ui.adsabs.harvard.edu/abs/2022arXiv220906831C} {p. arXiv:2209.06831}

\bibitem[\protect\citeauthoryear{{Chen}, {Avestruz}, {Kravtsov}, {Lau}  \&
  {Nagai}}{{Chen} et~al.}{2019}]{Chen2019}
{Chen} H.,  {Avestruz} C.,  {Kravtsov} A.~V.,  {Lau} E.~T.,   {Nagai} D.,
  2019, \mn@doi [\mnras] {10.1093/mnras/stz2776}, \href
  {https://ui.adsabs.harvard.edu/abs/2019MNRAS.490.2380C} {490, 2380}

\bibitem[\protect\citeauthoryear{{Churazov}, {Forman}, {Jones}, {Sunyaev}  \&
  {B{\"o}hringer}}{{Churazov} et~al.}{2004}]{Chu04}
{Churazov} E.,  {Forman} W.,  {Jones} C.,  {Sunyaev} R.,   {B{\"o}hringer} H.,
  2004, \mn@doi [\mnras] {10.1111/j.1365-2966.2004.07201.x}, \href
  {https://ui.adsabs.harvard.edu/abs/2004MNRAS.347...29C} {347, 29}

\bibitem[\protect\citeauthoryear{{Churazov} et~al.,}{{Churazov}
  et~al.}{2012}]{Chu12}
{Churazov} E.,  et~al., 2012, \mn@doi [\mnras]
  {10.1111/j.1365-2966.2011.20372.x}, \href
  {https://ui.adsabs.harvard.edu/abs/2012MNRAS.421.1123C} {421, 1123}

\bibitem[\protect\citeauthoryear{{De Luca}, {De Petris}, {Yepes}, {Cui},
  {Knebe}  \& {Rasia}}{{De Luca} et~al.}{2021}]{Del21}
{De Luca} F.,  {De Petris} M.,  {Yepes} G.,  {Cui} W.,  {Knebe} A.,   {Rasia}
  E.,  2021, \mn@doi [\mnras] {10.1093/mnras/stab1073}, \href
  {https://ui.adsabs.harvard.edu/abs/2021MNRAS.504.5383D} {504, 5383}

\bibitem[\protect\citeauthoryear{{Eckert}, {Gaspari}, {Vazza}, {Gastaldello},
  {Tramacere}, {Zimmer}, {Ettori}  \& {Paltani}}{{Eckert} et~al.}{2017}]{Eck17}
{Eckert} D.,  {Gaspari} M.,  {Vazza} F.,  {Gastaldello} F.,  {Tramacere} A.,
  {Zimmer} S.,  {Ettori} S.,   {Paltani} S.,  2017, \mn@doi [\apjl]
  {10.3847/2041-8213/aa7c1a}, \href
  {https://ui.adsabs.harvard.edu/abs/2017ApJ...843L..29E} {843, L29}

\bibitem[\protect\citeauthoryear{{Forman}, {Churazov}, {Jones}, {Heinz},
  {Kraft}  \& {Vikhlinin}}{{Forman} et~al.}{2017}]{Forman17}
{Forman} W.,  {Churazov} E.,  {Jones} C.,  {Heinz} S.,  {Kraft} R.,
  {Vikhlinin} A.,  2017, \mn@doi [\apj] {10.3847/1538-4357/aa70e4}, \href
  {https://ui.adsabs.harvard.edu/abs/2017ApJ...844..122F} {844, 122}

\bibitem[\protect\citeauthoryear{{Gaspari}, {Churazov}, {Nagai}, {Lau}  \&
  {Zhuravleva}}{{Gaspari} et~al.}{2014}]{Gas14}
{Gaspari} M.,  {Churazov} E.,  {Nagai} D.,  {Lau} E.~T.,   {Zhuravleva} I.,
  2014, \mn@doi [\aap] {10.1051/0004-6361/201424043}, \href
  {https://ui.adsabs.harvard.edu/abs/2014A&A...569A..67G} {569, A67}

\bibitem[\protect\citeauthoryear{{Gilfanov}, {Syunyaev}  \&
  {Churazov}}{{Gilfanov} et~al.}{1987}]{Gil87}
{Gilfanov} M.~R.,  {Syunyaev} R.~A.,   {Churazov} E.~M.,  1987, Soviet
  Astronomy Letters, \href
  {https://ui.adsabs.harvard.edu/abs/1987SvAL...13....3G} {13, 3}

\bibitem[\protect\citeauthoryear{{Gu} et~al.,}{{Gu} et~al.}{2009}]{Gu09}
{Gu} L.,  et~al., 2009, \mn@doi [\apj] {10.1088/0004-637X/700/2/1161}, \href
  {https://ui.adsabs.harvard.edu/abs/2009ApJ...700.1161G} {700, 1161}

\bibitem[\protect\citeauthoryear{{Harvey}, {Robertson}, {Tam}, {Jauzac},
  {Massey}, {Rhodes}  \& {McCarthy}}{{Harvey} et~al.}{2021}]{Har21}
{Harvey} D.,  {Robertson} A.,  {Tam} S.-I.,  {Jauzac} M.,  {Massey} R.,
  {Rhodes} J.,   {McCarthy} I.~G.,  2021, \mn@doi [\mnras]
  {10.1093/mnras/staa3193}, \href
  {https://ui.adsabs.harvard.edu/abs/2021MNRAS.500.2627H} {500, 2627}

\bibitem[\protect\citeauthoryear{{Hitomi Collaboration} et~al.,}{{Hitomi
  Collaboration} et~al.}{2018a}]{Hit18}
{Hitomi Collaboration} et~al., 2018a, \mn@doi [\pasj] {10.1093/pasj/psx138},
  \href {https://ui.adsabs.harvard.edu/abs/2018PASJ...70....9H} {70, 9}

\bibitem[\protect\citeauthoryear{{Hitomi Collaboration} et~al.,}{{Hitomi
  Collaboration} et~al.}{2018b}]{HitRS}
{Hitomi Collaboration} et~al., 2018b, \mn@doi [\pasj] {10.1093/pasj/psx127},
  \href {https://ui.adsabs.harvard.edu/abs/2018PASJ...70...10H} {70, 10}

\bibitem[\protect\citeauthoryear{{Hofmann}, {Sanders}, {Nandra}, {Clerc}  \&
  {Gaspari}}{{Hofmann} et~al.}{2016}]{Hof16}
{Hofmann} F.,  {Sanders} J.~S.,  {Nandra} K.,  {Clerc} N.,   {Gaspari} M.,
  2016, \mn@doi [\aap] {10.1051/0004-6361/201526925}, \href
  {https://ui.adsabs.harvard.edu/abs/2016A&A...585A.130H} {585, A130}

\bibitem[\protect\citeauthoryear{{Iapichino} \& {Niemeyer}}{{Iapichino} \&
  {Niemeyer}}{2008}]{Iap08}
{Iapichino} L.,  {Niemeyer} J.~C.,  2008, \mn@doi [\mnras]
  {10.1111/j.1365-2966.2008.13518.x}, \href
  {https://ui.adsabs.harvard.edu/abs/2008MNRAS.388.1089I} {388, 1089}

\bibitem[\protect\citeauthoryear{{Kawahara}, {Suto}, {Kitayama}, {Sasaki},
  {Shimizu}, {Rasia}  \& {Dolag}}{{Kawahara} et~al.}{2007}]{Kaw07}
{Kawahara} H.,  {Suto} Y.,  {Kitayama} T.,  {Sasaki} S.,  {Shimizu} M.,
  {Rasia} E.,   {Dolag} K.,  2007, \mn@doi [\apj] {10.1086/512231}, \href
  {https://ui.adsabs.harvard.edu/abs/2007ApJ...659..257K} {659, 257}

\bibitem[\protect\citeauthoryear{{Khatri} \& {Gaspari}}{{Khatri} \&
  {Gaspari}}{2016}]{Kha16}
{Khatri} R.,  {Gaspari} M.,  2016, \mn@doi [\mnras] {10.1093/mnras/stw2027},
  \href {https://ui.adsabs.harvard.edu/abs/2016MNRAS.463..655K} {463, 655}

\bibitem[\protect\citeauthoryear{{Kravtsov} \& {Klypin}}{{Kravtsov} \&
  {Klypin}}{1999}]{Kra99}
{Kravtsov} A.~V.,  {Klypin} A.~A.,  1999, \mn@doi [\apj] {10.1086/307495},
  \href {http://adsabs.harvard.edu/abs/1999ApJ...520..437K} {520, 437}

\bibitem[\protect\citeauthoryear{{Lau}, {Kravtsov}  \& {Nagai}}{{Lau}
  et~al.}{2009}]{Lau09}
{Lau} E.~T.,  {Kravtsov} A.~V.,   {Nagai} D.,  2009, \mn@doi [\apj]
  {10.1088/0004-637X/705/2/1129}, \href
  {https://ui.adsabs.harvard.edu/abs/2009ApJ...705.1129L} {705, 1129}

\bibitem[\protect\citeauthoryear{{Lau}, {Nagai}, {Kravtsov}, {Vikhlinin}  \&
  {Zentner}}{{Lau} et~al.}{2012}]{Lau12}
{Lau} E.~T.,  {Nagai} D.,  {Kravtsov} A.~V.,  {Vikhlinin} A.,   {Zentner}
  A.~R.,  2012, \mn@doi [\apj] {10.1088/0004-637X/755/2/116}, \href
  {https://ui.adsabs.harvard.edu/abs/2012ApJ...755..116L} {755, 116}

\bibitem[\protect\citeauthoryear{{Lovisari} et~al.,}{{Lovisari}
  et~al.}{2017}]{Lov17}
{Lovisari} L.,  et~al., 2017, \mn@doi [\apj] {10.3847/1538-4357/aa855f}, \href
  {https://ui.adsabs.harvard.edu/abs/2017ApJ...846...51L} {846, 51}

\bibitem[\protect\citeauthoryear{{Mantz}, {Allen}, {Morris}, {Schmidt}, {von
  der Linden}  \& {Urban}}{{Mantz} et~al.}{2015}]{Man15}
{Mantz} A.~B.,  {Allen} S.~W.,  {Morris} R.~G.,  {Schmidt} R.~W.,  {von der
  Linden} A.,   {Urban} O.,  2015, \mn@doi [\mnras] {10.1093/mnras/stv219},
  \href {https://ui.adsabs.harvard.edu/abs/2015MNRAS.449..199M} {449, 199}

\bibitem[\protect\citeauthoryear{{Miniati} \& {Beresnyak}}{{Miniati} \&
  {Beresnyak}}{2015}]{Min15}
{Miniati} F.,  {Beresnyak} A.,  2015, \mn@doi [\nat] {10.1038/nature14552},
  \href {https://ui.adsabs.harvard.edu/abs/2015Natur.523...59M} {523, 59}

\bibitem[\protect\citeauthoryear{{Mohapatra}, {Federrath}  \&
  {Sharma}}{{Mohapatra} et~al.}{2020}]{Moh20}
{Mohapatra} R.,  {Federrath} C.,   {Sharma} P.,  2020, \mn@doi [\mnras]
  {10.1093/mnras/staa711}, \href
  {https://ui.adsabs.harvard.edu/abs/2020MNRAS.493.5838M} {493, 5838}

\bibitem[\protect\citeauthoryear{{Mohapatra}, {Federrath}  \&
  {Sharma}}{{Mohapatra} et~al.}{2021}]{Moh21}
{Mohapatra} R.,  {Federrath} C.,   {Sharma} P.,  2021, \mn@doi [\mnras]
  {10.1093/mnras/staa3564}, \href
  {https://ui.adsabs.harvard.edu/abs/2021MNRAS.500.5072M} {500, 5072}

\bibitem[\protect\citeauthoryear{{Mohr}, {Evrard}, {Fabricant}  \&
  {Geller}}{{Mohr} et~al.}{1995}]{Moh95}
{Mohr} J.~J.,  {Evrard} A.~E.,  {Fabricant} D.~G.,   {Geller} M.~J.,  1995,
  \mn@doi [\apj] {10.1086/175852}, \href
  {https://ui.adsabs.harvard.edu/abs/1995ApJ...447....8M} {447, 8}

\bibitem[\protect\citeauthoryear{{Mroczkowski} et~al.,}{{Mroczkowski}
  et~al.}{2019}]{Mroczkowski2019}
{Mroczkowski} T.,  et~al., 2019, \mn@doi [\ssr] {10.1007/s11214-019-0581-2},
  \href {https://ui.adsabs.harvard.edu/abs/2019SSRv..215...17M} {215, 17}

\bibitem[\protect\citeauthoryear{{Nagai} \& {Lau}}{{Nagai} \&
  {Lau}}{2011}]{Nag11}
{Nagai} D.,  {Lau} E.~T.,  2011, \mn@doi [\apjl] {10.1088/2041-8205/731/1/L10},
  \href {https://ui.adsabs.harvard.edu/abs/2011ApJ...731L..10N} {731, L10}

\bibitem[\protect\citeauthoryear{{Nagai}, {Vikhlinin}  \& {Kravtsov}}{{Nagai}
  et~al.}{2007a}]{Nag07a}
{Nagai} D.,  {Vikhlinin} A.,   {Kravtsov} A.~V.,  2007a, \mn@doi [\apj]
  {10.1086/509868}, \href
  {https://ui.adsabs.harvard.edu/abs/2007ApJ...655...98N} {655, 98}

\bibitem[\protect\citeauthoryear{{Nagai}, {Kravtsov}  \& {Vikhlinin}}{{Nagai}
  et~al.}{2007b}]{Nag07b}
{Nagai} D.,  {Kravtsov} A.~V.,   {Vikhlinin} A.,  2007b, \mn@doi [\apj]
  {10.1086/521328}, \href
  {https://ui.adsabs.harvard.edu/abs/2007ApJ...668....1N} {668, 1}

\bibitem[\protect\citeauthoryear{{Nagai}, {Lau}, {Avestruz}, {Nelson}  \&
  {Rudd}}{{Nagai} et~al.}{2013}]{Nag13}
{Nagai} D.,  {Lau} E.~T.,  {Avestruz} C.,  {Nelson} K.,   {Rudd} D.~H.,  2013,
  \mn@doi [\apj] {10.1088/0004-637X/777/2/137}, \href
  {https://ui.adsabs.harvard.edu/abs/2013ApJ...777..137N} {777, 137}

\bibitem[\protect\citeauthoryear{{Nandra} et~al.,}{{Nandra}
  et~al.}{2013}]{athena13}
{Nandra} K.,  et~al., 2013, arXiv e-prints, \href
  {https://ui.adsabs.harvard.edu/abs/2013arXiv1306.2307N} {p. arXiv:1306.2307}

\bibitem[\protect\citeauthoryear{{Nelson}, {Rudd}, {Shaw}  \& {Nagai}}{{Nelson}
  et~al.}{2012}]{Nelson2012}
{Nelson} K.,  {Rudd} D.~H.,  {Shaw} L.,   {Nagai} D.,  2012, \mn@doi [\apj]
  {10.1088/0004-637X/751/2/121}, \href
  {https://ui.adsabs.harvard.edu/abs/2012ApJ...751..121N} {751, 121}

\bibitem[\protect\citeauthoryear{{Nelson}, {Lau}, {Nagai}, {Rudd}  \&
  {Yu}}{{Nelson} et~al.}{2014a}]{Nelson2014}
{Nelson} K.,  {Lau} E.~T.,  {Nagai} D.,  {Rudd} D.~H.,   {Yu} L.,  2014a,
  \mn@doi [\apj] {10.1088/0004-637X/782/2/107}, \href
  {https://ui.adsabs.harvard.edu/abs/2014ApJ...782..107N} {782, 107}

\bibitem[\protect\citeauthoryear{{Nelson}, {Lau}  \& {Nagai}}{{Nelson}
  et~al.}{2014b}]{Nelson2014b}
{Nelson} K.,  {Lau} E.~T.,   {Nagai} D.,  2014b, \mn@doi [\apj]
  {10.1088/0004-637X/792/1/25}, \href
  {https://ui.adsabs.harvard.edu/abs/2014ApJ...792...25N} {792, 25}

\bibitem[\protect\citeauthoryear{{Nulsen} et~al.,}{{Nulsen}
  et~al.}{2013}]{Nul13}
{Nulsen} P. E.~J.,  et~al., 2013, \mn@doi [\apj] {10.1088/0004-637X/775/2/117},
  \href {https://ui.adsabs.harvard.edu/abs/2013ApJ...775..117N} {775, 117}

\bibitem[\protect\citeauthoryear{{Nurgaliev}, {McDonald}, {Benson}, {Miller},
  {Stubbs}  \& {Vikhlinin}}{{Nurgaliev} et~al.}{2013}]{Nur13}
{Nurgaliev} D.,  {McDonald} M.,  {Benson} B.~A.,  {Miller} E.~D.,  {Stubbs}
  C.~W.,   {Vikhlinin} A.,  2013, \mn@doi [\apj] {10.1088/0004-637X/779/2/112},
  \href {https://ui.adsabs.harvard.edu/abs/2013ApJ...779..112N} {779, 112}

\bibitem[\protect\citeauthoryear{{Ogorzalek} et~al.,}{{Ogorzalek}
  et~al.}{2017}]{Ogo17}
{Ogorzalek} A.,  et~al., 2017, \mn@doi [\mnras] {10.1093/mnras/stx2030}, \href
  {https://ui.adsabs.harvard.edu/abs/2017MNRAS.472.1659O} {472, 1659}

\bibitem[\protect\citeauthoryear{{Ossenkopf}, {Krips}  \&
  {Stutzki}}{{Ossenkopf} et~al.}{2008}]{Oss08}
{Ossenkopf} V.,  {Krips} M.,   {Stutzki} J.,  2008, \mn@doi [\aap]
  {10.1051/0004-6361:20079106}, \href
  {https://ui.adsabs.harvard.edu/abs/2008A&A...485..917O} {485, 917}

\bibitem[\protect\citeauthoryear{{Pratt}, {Arnaud}, {Biviano}, {Eckert},
  {Ettori}, {Nagai}, {Okabe}  \& {Reiprich}}{{Pratt} et~al.}{2019}]{Pratt2019}
{Pratt} G.~W.,  {Arnaud} M.,  {Biviano} A.,  {Eckert} D.,  {Ettori} S.,
  {Nagai} D.,  {Okabe} N.,   {Reiprich} T.~H.,  2019, \mn@doi [\ssr]
  {10.1007/s11214-019-0591-0}, \href
  {https://ui.adsabs.harvard.edu/abs/2019SSRv..215...25P} {215, 25}

\bibitem[\protect\citeauthoryear{{Rasia} et~al.,}{{Rasia} et~al.}{2006}]{Ras06}
{Rasia} E.,  et~al., 2006, \mn@doi [\mnras] {10.1111/j.1365-2966.2006.10466.x},
  \href {https://ui.adsabs.harvard.edu/abs/2006MNRAS.369.2013R} {369, 2013}

\bibitem[\protect\citeauthoryear{{Rasia}, {Meneghetti}  \& {Ettori}}{{Rasia}
  et~al.}{2013}]{Ras13}
{Rasia} E.,  {Meneghetti} M.,   {Ettori} S.,  2013, \mn@doi [The Astronomical
  Review] {10.1080/21672857.2013.11519713}, \href
  {https://ui.adsabs.harvard.edu/abs/2013AstRv...8a..40R} {8, 40}

\bibitem[\protect\citeauthoryear{{Rasia} et~al.,}{{Rasia} et~al.}{2014}]{Ras14}
{Rasia} E.,  et~al., 2014, \mn@doi [\apj] {10.1088/0004-637X/791/2/96}, \href
  {https://ui.adsabs.harvard.edu/abs/2014ApJ...791...96R} {791, 96}

\bibitem[\protect\citeauthoryear{{Rebusco}, {Churazov}, {B{\"o}hringer}  \&
  {Forman}}{{Rebusco} et~al.}{2005}]{Reb05}
{Rebusco} P.,  {Churazov} E.,  {B{\"o}hringer} H.,   {Forman} W.,  2005,
  \mn@doi [\mnras] {10.1111/j.1365-2966.2005.08965.x}, \href
  {https://ui.adsabs.harvard.edu/abs/2005MNRAS.359.1041R} {359, 1041}

\bibitem[\protect\citeauthoryear{{Rudd}, {Zentner}  \& {Kravtsov}}{{Rudd}
  et~al.}{2008}]{Rudd08}
{Rudd} D.~H.,  {Zentner} A.~R.,   {Kravtsov} A.~V.,  2008, \mn@doi [\apj]
  {10.1086/523836}, \href {http://adsabs.harvard.edu/abs/2008ApJ...672...19R}
  {672, 19}

\bibitem[\protect\citeauthoryear{{Sanders}, {Fabian}, {Smith}  \&
  {Peterson}}{{Sanders} et~al.}{2010}]{San10}
{Sanders} J.~S.,  {Fabian} A.~C.,  {Smith} R.~K.,   {Peterson} J.~R.,  2010,
  \mn@doi [\mnras] {10.1111/j.1745-3933.2009.00789.x}, \href
  {https://ui.adsabs.harvard.edu/abs/2010MNRAS.402L..11S} {402, L11}

\bibitem[\protect\citeauthoryear{{Sanders} et~al.,}{{Sanders}
  et~al.}{2020}]{San20}
{Sanders} J.~S.,  et~al., 2020, \mn@doi [\aap] {10.1051/0004-6361/201936468},
  \href {https://ui.adsabs.harvard.edu/abs/2020A&A...633A..42S} {633, A42}

\bibitem[\protect\citeauthoryear{{Santos}, {Rosati}, {Tozzi}, {B{\"o}hringer},
  {Ettori}  \& {Bignamini}}{{Santos} et~al.}{2008}]{San08}
{Santos} J.~S.,  {Rosati} P.,  {Tozzi} P.,  {B{\"o}hringer} H.,  {Ettori} S.,
  {Bignamini} A.,  2008, \mn@doi [\aap] {10.1051/0004-6361:20078815}, \href
  {https://ui.adsabs.harvard.edu/abs/2008A&A...483...35S} {483, 35}

\bibitem[\protect\citeauthoryear{{Schekochihin} \& {Cowley}}{{Schekochihin} \&
  {Cowley}}{2006}]{Sch06}
{Schekochihin} A.~A.,  {Cowley} S.~C.,  2006, \mn@doi [Physics of Plasmas]
  {10.1063/1.2179053}, \href
  {https://ui.adsabs.harvard.edu/abs/2006PhPl...13e6501S} {13, 056501}

\bibitem[\protect\citeauthoryear{{Schuecker}, {Finoguenov}, {Miniati},
  {B{\"o}hringer}  \& {Briel}}{{Schuecker} et~al.}{2004}]{Sch04}
{Schuecker} P.,  {Finoguenov} A.,  {Miniati} F.,  {B{\"o}hringer} H.,   {Briel}
  U.~G.,  2004, \mn@doi [\aap] {10.1051/0004-6361:20041039}, \href
  {https://ui.adsabs.harvard.edu/abs/2004A&A...426..387S} {426, 387}

\bibitem[\protect\citeauthoryear{{Shi} \& {Zhang}}{{Shi} \&
  {Zhang}}{2019}]{Shi19}
{Shi} X.,  {Zhang} C.,  2019, \mn@doi [\mnras] {10.1093/mnras/stz1392}, \href
  {https://ui.adsabs.harvard.edu/abs/2019MNRAS.487.1072S} {487, 1072}

\bibitem[\protect\citeauthoryear{{Shi}, {Komatsu}, {Nagai}  \& {Lau}}{{Shi}
  et~al.}{2016}]{Shi16}
{Shi} X.,  {Komatsu} E.,  {Nagai} D.,   {Lau} E.~T.,  2016, \mn@doi [\mnras]
  {10.1093/mnras/stv2504}, \href
  {https://ui.adsabs.harvard.edu/abs/2016MNRAS.455.2936S} {455, 2936}

\bibitem[\protect\citeauthoryear{{Shi}, {Nagai}  \& {Lau}}{{Shi}
  et~al.}{2018}]{Shi18}
{Shi} X.,  {Nagai} D.,   {Lau} E.~T.,  2018, \mn@doi [\mnras]
  {10.1093/mnras/sty2340}, \href
  {https://ui.adsabs.harvard.edu/abs/2018MNRAS.481.1075S} {481, 1075}

\bibitem[\protect\citeauthoryear{{Shi}, {Nagai}, {Aung}  \& {Wetzel}}{{Shi}
  et~al.}{2020}]{Shi20}
{Shi} X.,  {Nagai} D.,  {Aung} H.,   {Wetzel} A.,  2020, \mn@doi [\mnras]
  {10.1093/mnras/staa1221}, \href
  {https://ui.adsabs.harvard.edu/abs/2020MNRAS.495..784S} {495, 784}

\bibitem[\protect\citeauthoryear{{Simonte}, {Vazza}, {Brighenti},
  {Br{\"u}ggen}, {Jones}  \& {Angelinelli}}{{Simonte} et~al.}{2022}]{Sim22}
{Simonte} M.,  {Vazza} F.,  {Brighenti} F.,  {Br{\"u}ggen} M.,  {Jones} T.~W.,
   {Angelinelli} M.,  2022, \mn@doi [\aap] {10.1051/0004-6361/202141703}, \href
  {https://ui.adsabs.harvard.edu/abs/2022A&A...658A.149S} {658, A149}

\bibitem[\protect\citeauthoryear{{Tamura} et~al.,}{{Tamura}
  et~al.}{2014}]{Tam14}
{Tamura} T.,  et~al., 2014, \mn@doi [\apj] {10.1088/0004-637X/782/1/38}, \href
  {https://ui.adsabs.harvard.edu/abs/2014ApJ...782...38T} {782, 38}

\bibitem[\protect\citeauthoryear{{Vazza}, {Brunetti}, {Kritsuk}, {Wagner},
  {Gheller}  \& {Norman}}{{Vazza} et~al.}{2009}]{Vaz09}
{Vazza} F.,  {Brunetti} G.,  {Kritsuk} A.,  {Wagner} R.,  {Gheller} C.,
  {Norman} M.,  2009, \mn@doi [\aap] {10.1051/0004-6361/200912535}, \href
  {https://ui.adsabs.harvard.edu/abs/2009A&A...504...33V} {504, 33}

\bibitem[\protect\citeauthoryear{{Vazza}, {Jones}, {Br{\"u}ggen}, {Brunetti},
  {Gheller}, {Porter}  \& {Ryu}}{{Vazza} et~al.}{2017}]{Vaz17}
{Vazza} F.,  {Jones} T.~W.,  {Br{\"u}ggen} M.,  {Brunetti} G.,  {Gheller} C.,
  {Porter} D.,   {Ryu} D.,  2017, \mn@doi [\mnras] {10.1093/mnras/stw2351},
  \href {https://ui.adsabs.harvard.edu/abs/2017MNRAS.464..210V} {464, 210}

\bibitem[\protect\citeauthoryear{{Walker}, {Sanders}  \& {Fabian}}{{Walker}
  et~al.}{2015}]{Wal15}
{Walker} S.~A.,  {Sanders} J.~S.,   {Fabian} A.~C.,  2015, \mn@doi [\mnras]
  {10.1093/mnras/stv1929}, \href
  {https://ui.adsabs.harvard.edu/abs/2015MNRAS.453.3699W} {453, 3699}

\bibitem[\protect\citeauthoryear{{Wei{\ss}mann}, {B{\"o}hringer},
  {{\v{S}}uhada}  \& {Ameglio}}{{Wei{\ss}mann} et~al.}{2013}]{We13}
{Wei{\ss}mann} A.,  {B{\"o}hringer} H.,  {{\v{S}}uhada} R.,   {Ameglio} S.,
  2013, \mn@doi [\aap] {10.1051/0004-6361/201219333}, \href
  {https://ui.adsabs.harvard.edu/abs/2013A&A...549A..19W} {549, A19}

\bibitem[\protect\citeauthoryear{{XRISM Science Team}}{{XRISM Science
  Team}}{2020}]{xrism20}
{XRISM Science Team} 2020, arXiv e-prints, \href
  {https://ui.adsabs.harvard.edu/abs/2020arXiv200304962X} {p. arXiv:2003.04962}

\bibitem[\protect\citeauthoryear{{Yuan} \& {Han}}{{Yuan} \&
  {Han}}{2020}]{Yua20}
{Yuan} Z.~S.,  {Han} J.~L.,  2020, \mn@doi [\mnras] {10.1093/mnras/staa2363},
  \href {https://ui.adsabs.harvard.edu/abs/2020MNRAS.497.5485Y} {497, 5485}

\bibitem[\protect\citeauthoryear{{Zemp}, {Gnedin}, {Gnedin}  \&
  {Kravtsov}}{{Zemp} et~al.}{2011}]{Zemp2011}
{Zemp} M.,  {Gnedin} O.~Y.,  {Gnedin} N.~Y.,   {Kravtsov} A.~V.,  2011, \mn@doi
  [\apjs] {10.1088/0067-0049/197/2/30}, \href
  {https://ui.adsabs.harvard.edu/abs/2011ApJS..197...30Z} {197, 30}

\bibitem[\protect\citeauthoryear{{Zhuravleva}, {Churazov}, {Kravtsov}, {Lau},
  {Nagai}  \& {Sunyaev}}{{Zhuravleva} et~al.}{2013}]{Zhuravleva2013}
{Zhuravleva} I.,  {Churazov} E.,  {Kravtsov} A.,  {Lau} E.~T.,  {Nagai} D.,
  {Sunyaev} R.,  2013, \mn@doi [\mnras] {10.1093/mnras/sts275}, \href
  {https://ui.adsabs.harvard.edu/abs/2013MNRAS.428.3274Z} {428, 3274}

\bibitem[\protect\citeauthoryear{{Zhuravleva} et~al.,}{{Zhuravleva}
  et~al.}{2014a}]{Zhu14b}
{Zhuravleva} I.,  et~al., 2014a, \mn@doi [\nat] {10.1038/nature13830}, \href
  {https://ui.adsabs.harvard.edu/abs/2014Natur.515...85Z} {515, 85}

\bibitem[\protect\citeauthoryear{{Zhuravleva} et~al.,}{{Zhuravleva}
  et~al.}{2014b}]{Zhuravleva2014}
{Zhuravleva} I.,  et~al., 2014b, \mn@doi [\apjl] {10.1088/2041-8205/788/1/L13},
  \href {https://ui.adsabs.harvard.edu/abs/2014ApJ...788L..13Z} {788, L13}

\bibitem[\protect\citeauthoryear{{Zhuravleva} et~al.,}{{Zhuravleva}
  et~al.}{2015}]{Zhu15}
{Zhuravleva} I.,  et~al., 2015, \mn@doi [\mnras] {10.1093/mnras/stv900}, \href
  {https://ui.adsabs.harvard.edu/abs/2015MNRAS.450.4184Z} {450, 4184}

\bibitem[\protect\citeauthoryear{{Zhuravleva}, {Allen}, {Mantz}  \&
  {Werner}}{{Zhuravleva} et~al.}{2018}]{Zhu18}
{Zhuravleva} I.,  {Allen} S.~W.,  {Mantz} A.,   {Werner} N.,  2018, \mn@doi
  [\apj] {10.3847/1538-4357/aadae3}, \href
  {https://ui.adsabs.harvard.edu/abs/2018ApJ...865...53Z} {865, 53}

\makeatother
\end{thebibliography}




\appendix
\label{app:A}
\section{$M_{\rm 1d} - \delta\xi/\xi$ correlations after removing radial variations}

\begin{figure*}
    \centering
    \includegraphics[width=\linewidth, trim=0 0 0 0]{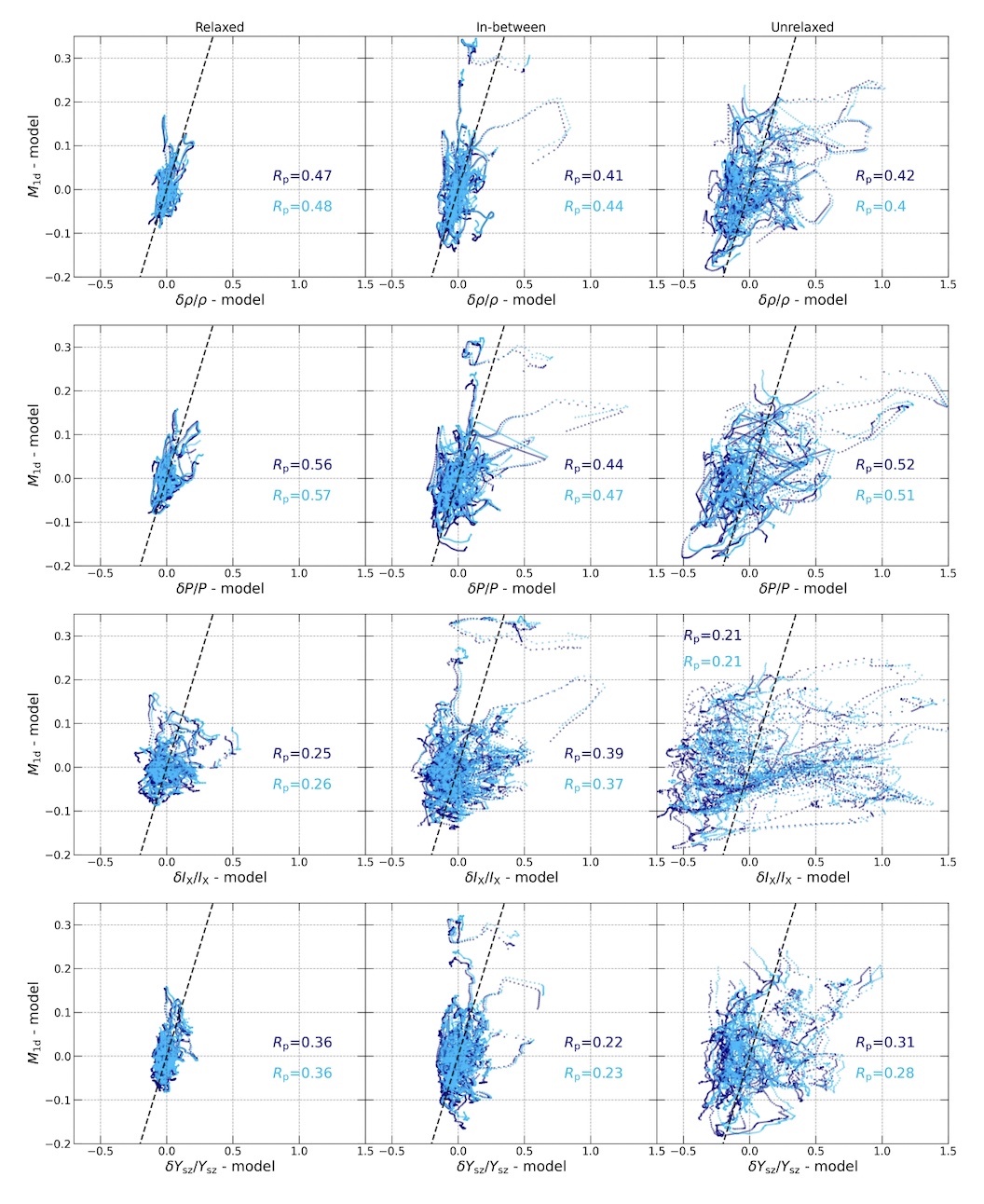}
    \caption{The same correlations as in Fig. \ref{fig:corr_all} with the radial variations of amplitudes and Mach numbers being removed by subtracting corresponding best-fitting linear (navy) or power-law (light blue) models. The corresponding Pearson ranks are shown in legends. The dashed lines indicate a one-to-one match for visual guidance.}
    \label{fig:corr_nor}
\end{figure*}

It is interesting to check a residual correlation between the Mach number of gas motions and fluctuation amplitudes after removing the radial variations of these characteristics. Fig. \ref{fig:corr_nor} shows the distributions after the best-fitting radial trends were subtracted from the initial $M_{\rm 1d}$ (see Table \ref{tab:m1d_correlations} for the best-fitting parameters) and $\delta\xi/\xi$. Both linear (navy) and power-law (light blue) models for the radial trends were considered. One can see that (1) the correlations with 3D amplitudes remain moderately strong with $R_{\rm p}\sim 0.4-0.6$ for clusters in all dynamic states; 
(2) the scatter is the smallest for relaxed clusters; (3) projected amplitudes show a weak correlation with Mach ($R_{\rm p} <0.4$), especially in dynamically-perturbed clusters. Despite smaller values of $R_{\rm p}$ compared to those in Fig. \ref{fig:corr_all}, this result indicates the fundamental origin of the correlation between the velocities of gas motions and gas fluctuations in the ICM.

\bsp	
\label{lastpage}
\end{document}